\documentclass[conference]{IEEEtran}
\IEEEoverridecommandlockouts
\usepackage{cite}
\usepackage{amsmath,amssymb,amsfonts}
\usepackage{algorithmic}
\usepackage{graphicx}
\usepackage{textcomp}
\usepackage{xcolor}
\usepackage{caption}
\usepackage{subcaption}
\def\BibTeX{{\rm B\kern-.05em{\sc i\kern-.025em b}\kern-.08em
    T\kern-.1667em\lower.7ex\hbox{E}\kern-.125emX}}
\begin{document}

\title{Is Disaggregation possible for HPC Cognitive Simulation?
}

\author{
    \IEEEauthorblockN{Michael R. Wyatt II\IEEEauthorrefmark{1}, Valen
      Yamamoto\IEEEauthorrefmark{1}, Zo\"{e} Tosi\IEEEauthorrefmark{1}, Ian
      Karlin\IEEEauthorrefmark{2}, Brian Van Essen\IEEEauthorrefmark{1}}
    \\\IEEEauthorblockA{\IEEEauthorrefmark{1}Center for Applied Scientific Computing\\
    Lawrence Livermore National Laboratory
    \\\{wyatt5, yamamoto6, tosi1, vanessen1\}@llnl.gov}
    \IEEEauthorblockA{\IEEEauthorrefmark{2}Intel Corporation
    \\\{ian.karlin\}@intel.com}
}

\maketitle

\begin{abstract}
Cognitive simulation (CogSim) is an important and emerging workflow for HPC
scientific exploration and scientific machine learning (SciML).   One
challenging workload for CogSim is the replacement of one component in a
complex physical simulation with a fast, learned, surrogate model that is
``inside'' of the computational loop.  The execution of this
\texttt{in-the-loop} inference is particularly challenging because it requires
frequent inference across multiple possible target models, can be on the
simulation's critical path (latency bound), is subject to requests from
multiple MPI ranks, and typically contains a small number of samples per
request.  In this paper we explore the use of large, dedicated Deep Learning /
AI accelerators that are disaggregated from compute nodes for this CogSim
workload.  We compare the trade-offs of using these accelerators versus the
node-local GPU accelerators on leadership-class HPC systems.
\end{abstract}

\begin{IEEEkeywords}
deep learning, cognitive simulation, surrogate models, accelerator performance
\end{IEEEkeywords}

\section{Introduction}

Recent advances in deep learning are leading to a resurgence of interest in the
use of surrogate or reduced-order models to replace complex, first-principle
physics packages in scientific simulations.  Rather than expecting data-driven
models to provide end-to-end predictions that replace entire multi-physics
simulations, we explore new hybrid workflows that intertwine data-driven,
learned models with traditional scientific simulation, a pattern in the
community called Cognitive Simulation (CogSim).  In this work, we explore
the use of discrete, disaggregated accelerator hardware coupled with leadership
class HPC resources to accelerate these workflows.

CogSim applications leverage AI for surrogate modeling of costly physics
calculations~\cite{Kluth_2020, Humbird_2021, Anirudh9741, Sun_2020}, and are
being developed to improve searching in multi-dimensional spaces and to
automate manual processes.  While the AI portion of the workflow is important
and consumes significant compute resources, traditional scientific simulation
CogSim workflows still require significant HPC compute capabilities -- often at
higher numerical precision.

Most of today's leadership-class HPC systems use GPU accelerators for both the
HPC and AI portions of CogSim workflows.  While many research efforts are
looking into other accelerators for HPC compute, no leading contenders have
emerged.  However, a large number of startup companies and traditional chip
vendors are producing AI optimized accelerators, up to and including dedicated
self-hosted machines.  These machines show promising speedups for a
variety of machine learning workloads~\cite{9387491}.

With different processor architectures optimal for different pieces of a CogSim
workload, an opportunity exists to build more cost efficient machines through
the addition of another accelerator type.  Accelerators could be integrated
into every node or they can be treated as a disaggregated resource by being
placed on the high-speed network of a supercomputer.  Each approach has
different cost and performance characteristics and the best approach may be
workload dependent.

Using the newly installed SambaNova DataScale\textregistered~system at Lawrence
Livermore National Laboratory (LLNL) as motivation, we explore the question of
whether the disaggregated design approach is viable for two test applications:
a non-local thermodynamic equilibrium (NLTE) collisional-radiative atomic physics
package~\cite{Kluth_2020} and a materials interface reconstruction task.  We
look at high-frequency machine learning inference of surrogate models due to
the challenges this use case presents in time to solution and data transfer
needs, and we aim to identify the most effective methods for performing it on each
system.  We aim to answer the question of should Deep Learning (DL) inference
be performed on the GPUs, competing with the primary calculations, or could
they be offloaded to dedicated AI accelerators, where the DL
calculation could occur simultaneously as the physics progresses on the GPU.

In this paper we make the following novel contributions:

\begin{itemize}
  \item We describe a first of its kind Disaggregated CogSim system and how it
        was built.  We also discuss how CogSim applications can use the disaggregated system and
        the relative challenges and costs.
  \item We show the trade-offs in terms of latency and throughput for processing
        inference samples on GPUs and AI accelerators for two independent
        surrogate models.
\end{itemize}

Our initial experimentation shows that disaggregated AI accelerators are viable
from a technical perspective to speed-up surrogate inference calculations that
occur in the Hydra multi-physics code used for simulating experiments on the
National Ignition Facility at LLNL.  Furthermore, our experimentation with a
material interface reconstruction surrogate demonstrates some of the challenges
with both developing these models and mapping them to emerging accelerators.

\section{Disaggregated Heterogeneous System Architectures for CogSim}
\label{sec:background}

CogSim is an emerging field that is generating new workflows and system
requirements.  Combining traditional HPC with ML to solve scientific problems,
CogSim applications have unique requirements from traditional HPC applications.
In this section, we describe one disaggregated system we have built for CogSim
workloads and discuss how various CogSim use cases can benefit.

\subsection{CogSim Disaggregated System Architectures}

Disaggregated system architectures with heterogeneous node types have multiple
advantages.  By offering disparate node types, different parts of a calculation
can execute on the most efficient compute resource.  In addition, these architectures allow each
application to run on the right mix of node types for its compute needs. This
solves the stranded resource problem that arises when a system is built with a
single heterogeneous node type with fixed ratios and multiple accelerators,
where accelerator or CPU resources may be wasted.  Disaggregation comes with
costs though as network attached resources are connected with lower bandwidth
and higher latency than node-local accelerators.

CogSim is a natural place to test whether disaggregated heterogeneous system
architectures could work in practice.  New accelerators show significant
performance gains for ML applications.  In addition, neural network training
and inference calculations have small input and output requirements relative to
their compute needs. Therefore, the impact of reduced bandwidth and additional
latency from network attached resources should be minimal with this workload.

Using the DataScale
AI accelerator from SambaNova Systems purchased by LLNL, we tested its ability
to serve as a disaggregated CogSim system.
This accelerator was integrated into
the high speed network of the Corona supercomputer, which has a theoretical
peak of over 11 PF.  The DataScale node is attached to a top of the rack (TOR)
switch that connects into the Corona core switches.  Communication between
Corona compute nodes and the DataScale system happens across a Mellanox
Infiniband ConnectX-6 with up to 100Gb/s bandwidth and less than $1\mu$s
latency.

The DataScale system houses 8 SambaNova Reconfigurable Dataflow Units
(RDU)\texttrademark~that utilize the SambaNova Reconfigurable Dataflow
Architecture (RDA)\texttrademark. DL models are compiled and run on the system
using the SambaNova SambaFlow\texttrademark~software stack.  Scheduling on the
DataScale system is done with the SLURM job manager.

\subsection{CogSim Workloads on Disaggregated Machines}
\label{sec:cogsim}

\begin{figure}[tbh]
\centering
  \includegraphics[width=1.0\columnwidth]{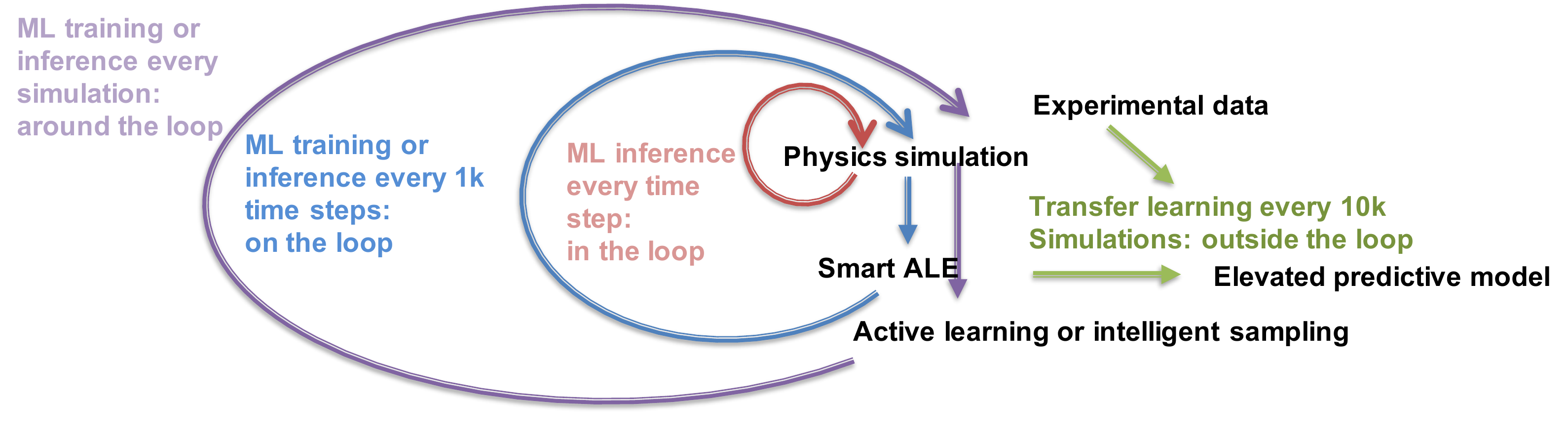}
\caption{Various uses of CogSim within a scientific simulation.}
\label{fig:cogsim_loops}
\end{figure}

Within the field of CogSim there are multiple ways that AI is incorporated into
scientific computations.  Figure \ref{fig:cogsim_loops} shows examples of three
``classes'' of possible use cases and how they are embedded within the simulation:
\texttt{in-the-loop}, \texttt{on-the-loop}, and \texttt{around-the-loop}.  Starting
from the center, \texttt{in-the-loop} represent some of the most challenging
tasks for integration because AI models are part of the inner most loop and
thus typically on the critical path.  Tasks can range from inference calculations of surrogate models
that replace physics computations that occur on every simulated zone for every timestep,
to inference calculations that help reconstruct mesh zones on material
boundaries.  These \texttt{in-the-loop} problems typically work with small batch
sizes and should be tuned to minimize inference latency.  Additionally, it is
quite possible to have multiple models for \texttt{in-the-loop} inference that
are used in different physics regimes, and thus should support concurrent execution.

The \texttt{around-the-loop} training problem presents the easiest case for adapting
to a disaggregated system.  Outer loop training takes inputs from one or more
simulations.  Since all of the data needs to be aggregated in one place for
training and training is more compute intensive than inference, the relative
network bandwidth and latency requirements are relaxed compared to the other
use cases that we present.  The main challenge for outer loop training is balancing
resource types and making sure models are retrained fast enough to improve the
quality of the workflow as new data are produced.

Presenting more of a challenge is training/retraining during a simulation or
inference that occurs \texttt{on-the-loop} every $x$ timesteps.  While more tightly
coupled, updating these models is not urgent and retraining of the model will
often include old data that is best stored elsewhere and not on the memory-constrained
HPC compute nodes. \texttt{On-the-loop} inference problems often permit
significant time for the computation to occur before an inference result is needed.

The two outermost use cases result in system
balance and allocation challenges to avoid stranded resources due to bursty use.
They also present bandwidth and latency requirements,
but those are not the strictest requirements.
Of the three use cases, the most tightly coupled physics surrogate models that run
\texttt{in-the-loop} present the largest challenge to using a disaggregated system.
This use case sends significant data per simulated zone to the
accelerator, and it has low latency tolerance for a response.  Due to these
challenges we focus on such \texttt{in-the-loop} calculations in this paper.  We
believe that if performing these calculations are viable using a disaggregated
accelerator, then it follows that the other two use cases will be as well.

\section{Related Work}
In this paper, we focus on many new and emerging technologies and techniques,
like CogSim and dataflow architectures. There are a small number of prior
publications we can look to for context of our work. For example,
in~\cite{9387491} the SambaNova Reconfigurable DataFlow is compared to
an Nvidia V100 GPU. The authors focus on the ability to train larger models on
the DataScale that would typically not fit on GPU. They demonstrate how larger
networks and input data sizes on dataflow architectures generate more accurate
models than the GPU in the same number of epochs. This work is largely focused on
typical ML workloads, such as image classification and NLP tasks.

A number of publications demonstrate the use of surrogate modeling in physics
applications~\cite{Sun_2020,Anirudh9741,Humbird_2021}. These works introduce
the use of surrogate models for \texttt{in-the-loop} inference with physics
applications like inertial confinement fusion~\cite{Humbird_2021,Anirudh9741}
and fluid simulation~\cite{Sun_2020}. While these publications focus on how the
physics applications can be performance optimized through CogSim, they do not
explore how different architectures and configurations (e.g., remote vs local
inference) can be adapted to the CogSim workload.  In this paper, we extend
the current scope of the literature by focusing on how traditional GPU
architectures and newer dataflow architectures perform with CogSim workloads.

In extending the CogSim and dataflow architecture literature, we also look to
many publications related to Deep Learning GPU performance. For example,
in~\cite{MITTAL2019101635} the authors survey the scope of recent publications that
demonstrate optimizing DL tasks on GPUs. Similar to publications outlined
in~\cite{MITTAL2019101635}, we test different APIs, toolkits, software,
mini-batch sizes, and hardware to optimize the performance on both GPU and
dataflow architectures. More specifically, we use
PyTorch~\cite{paszke2019pytorch} and focus on inference
performance~\cite{holmes2019grnn,zlateski2016znni,xu2018deep}. We test and compare
neural network inference performance on both Nvidia and AMD
GPUs~\cite{doi:10.1177/10943420211008288}. For each GPU architecture, we
compare multiple generations of GPU~\cite{10.1145/3184407.3184423}. We also
test toolkits and APIs like TensorRT~\cite{vanholder2016efficient} for
improving inference performance on Nvidia GPUs~\cite{8641600,xu2018deep}.

From these works, we gather best practices that we applied in our own
performance experiments. For example, one way that we ensure a fair comparison
between architectures in the context of CogSim workloads is considering data
movement and if it should be included in latency and throughput
measurements~\cite{5762730}. Unlike these previous publications, we extend the
performance measurement work to dataflow architectures. In doing so, we
establish novelty of our work in the neural network inference performance and CogSim space.

\section{In-the-loop inference}
\label{sec:models}

As discussed in Section \ref{sec:cogsim}, one of the most challenging tasks for
a disaggregated AI accelerator is to support \texttt{in-the-loop} inference.
The challenges are that the inference is typically on the critical path of the
primary scientific calculation, may require multiple independent models to serve
different inference tasks, will typically have small numbers of requests per
time-step, and will have to serve requests from multiple compute nodes (and
distributed MPI ranks).  These system integration challenges are why this paper
focuses on two data-driven surrogate models: Hermit and
MIR.  It is also important to note that neither these models are massively large,
or utilize complex neural network architectures or layers, and as a result allow
this study to focus on the system level issues.

\subsection{Hermit: a surrogate model for NLTE collisional-radiative atomic physics}

\begin{figure*}[!tbh]
\begin{subfigure}{0.33\textwidth}
\centering
  \includegraphics[width=\columnwidth]{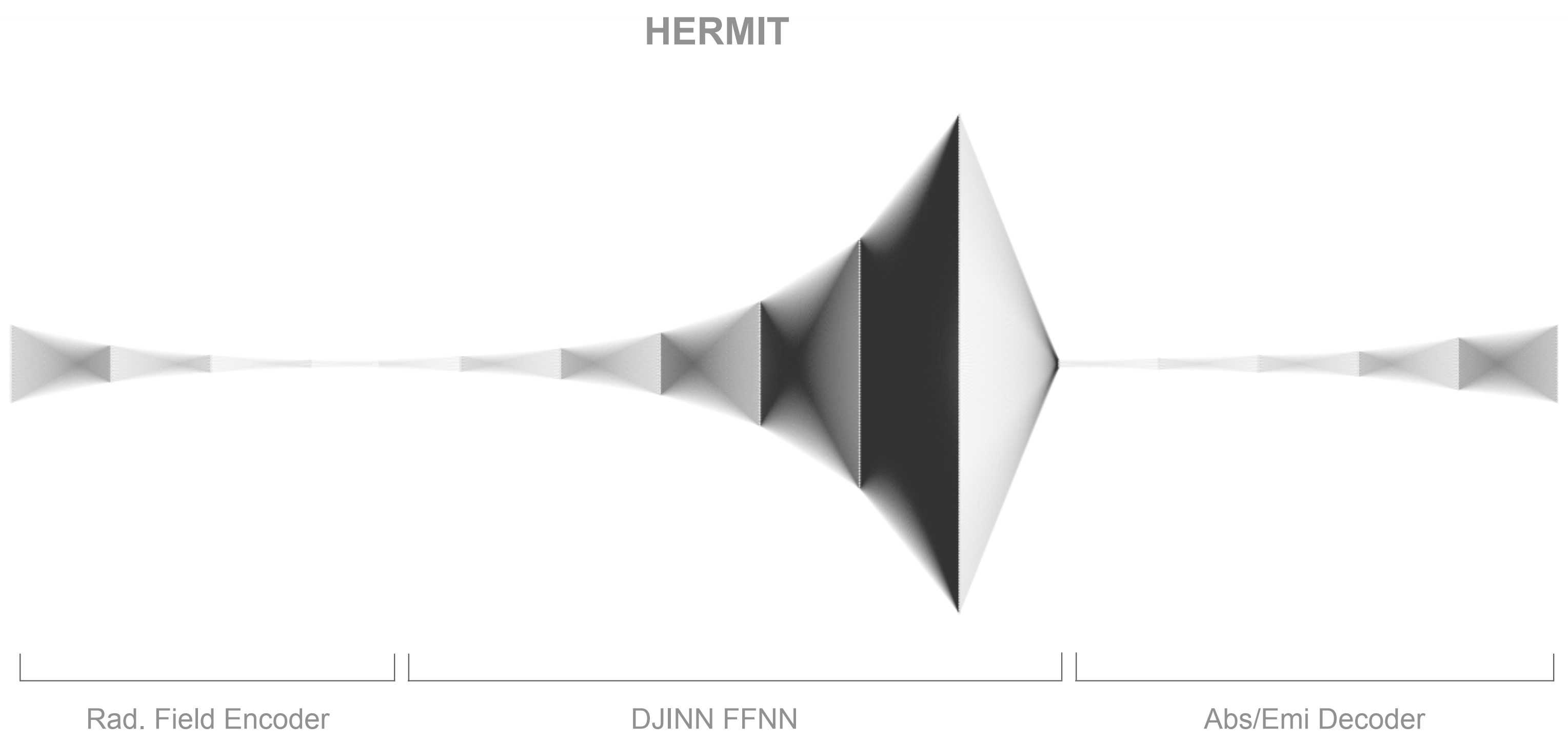}
\caption{Hermit neural network structure}
\label{fig:hermit}
\end{subfigure}
\begin{subfigure}{0.33\textwidth}
\centering
  \includegraphics[width=0.95\columnwidth]{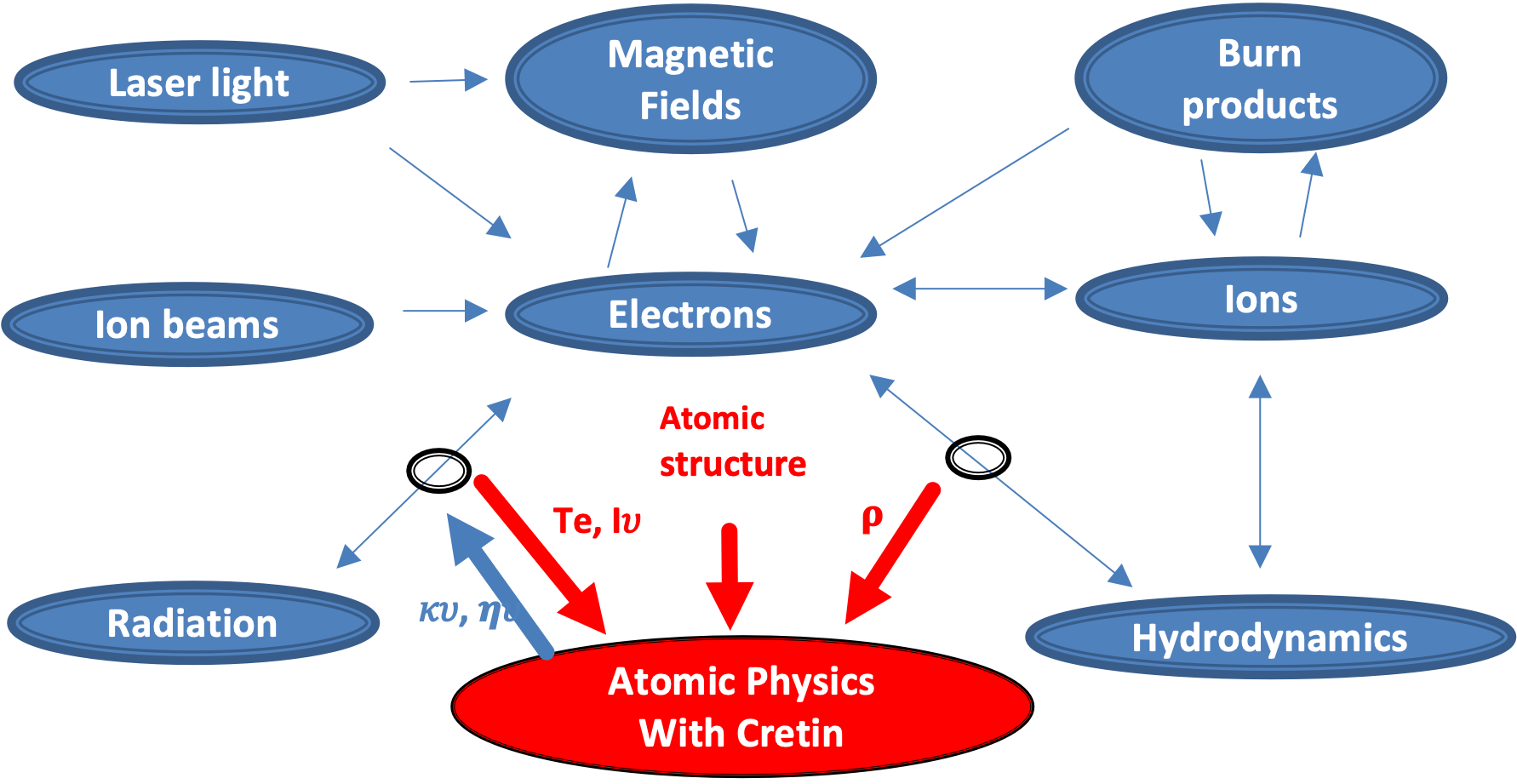}
\caption{Hydrodynamics with DCA physics}
\label{fig:hydra_cretin}
\end{subfigure}
\begin{subfigure}{0.33\textwidth}
\centering
  \includegraphics[width=0.95\columnwidth]{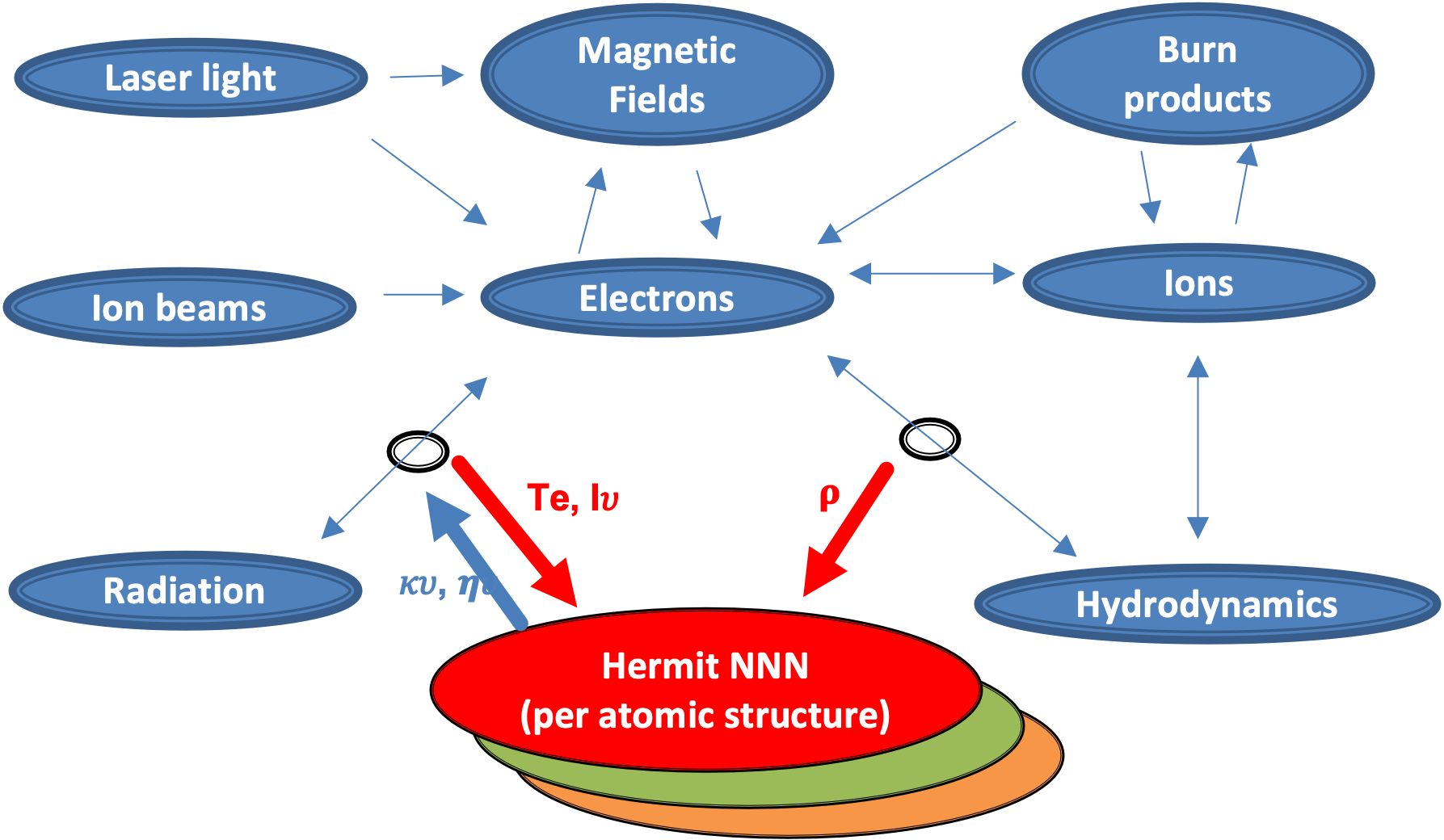}
\caption{Replacing DCA with Hermit model(s)\quad (one per atomic structure)}
\label{fig:hydra_hermit}
\end{subfigure}
\caption{Hermit model and interaction with hydrodynamics multi-physics simulation.}
\end{figure*}

The Hermit neural network model described in \cite{Kluth_2020} 
is diagrammed in Figure~\ref{fig:hermit}. The
model consists of 21 fully connected layers across 3 sub-structures: an
encoder, Deep Jointly-Informed Neural Network (DJINN) layers, and a decoder.
The total input size for each sample is just 42 values. The encoder has 4
layers with a maximum hidden layer width of 19 and the decoder has 6 layers
with a maximum hidden layer width of 27 neurons. The bulk of the network size
from the DJINN layers, which reach a maximum width of 2050 neurons. In total,
there are $2.8M$ parameters in the Hermit model.

We use the description of the Hermit model integration with the Hydra physics
simulation code in~\cite{doi:10.1063/1.872004} to characterize the needs of our
\texttt{in-the-loop} inference requirements and test our hypothesis that
disaggregated systems are feasible for CogSim applications.  Hermit is used to
replace the Detailed Configuration Accounting (DCA) package and requires two or
three inference calculations per zone in each timestep of the simulation.
Typical simulations can be tens of thousands of timesteps. The relationship
between DCA or Hermit and the rest if Hydra is shown in Figure
\ref{fig:hydra_cretin} and \ref{fig:hydra_hermit}.

While problem and use case dependent, typical Hydra problems using DCA physics
are run with only a few zones per node due to high memory capacity needs and/or
time to solution requirements~\cite{10.1007/978-3-319-17353-5_15}.
On Nvidia V100 GPUs, users typically run 100-1,000 zones
per GPU when using DCA physics models.
Swapping in the Hermit model both accelerates the calculation and
reduces the memory footprint, enabling the user to run with more zones per GPU.
In Hydra, two or three inference calculations are required per zone in each
timestep of the simulation.
With 10,000 zones per GPU, 20,000 - 30,000 inference calculations are needed per timestep.
In addition, inference requests from each MPI rank are submitted to different Hermit models,
where each model is trained to represent a particular material.
An MPI rank might typically require results for 5-10 different materials.
The low number of inference calculations needed and the fact that they are
spread across multiple models means small batch size performance is key to
Hermit performance.

\subsection{Material Interface Reconstruction (MIR) Model}

\begin{figure*}[!tbh]
\begin{subfigure}{0.4\textwidth}
\centering
 \includegraphics[width=\columnwidth]{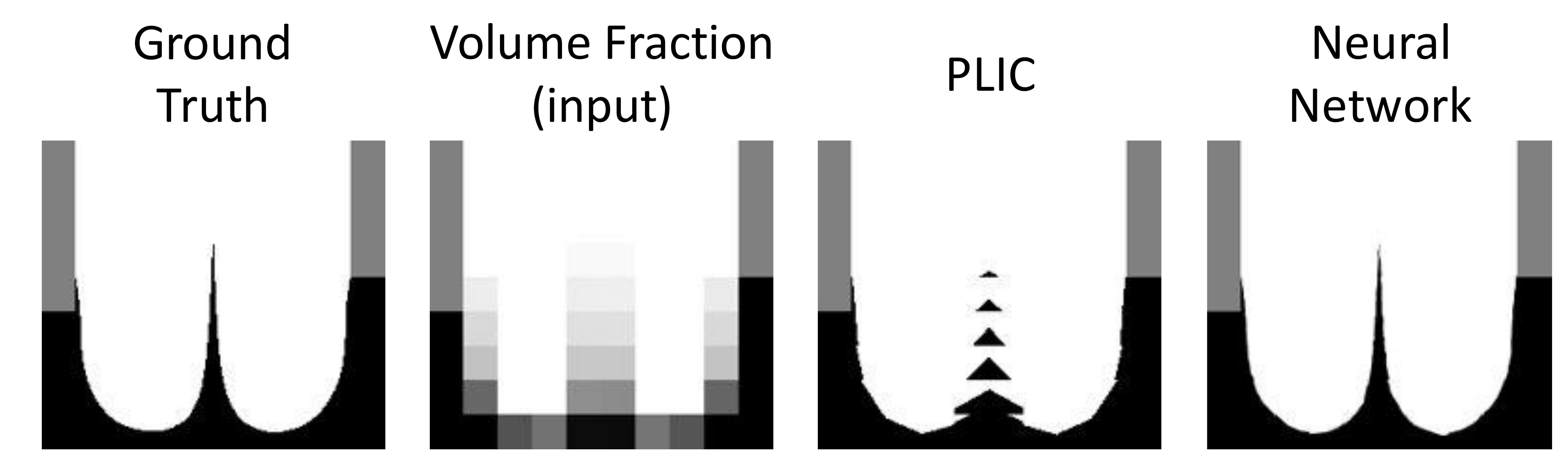}
 \caption{Comparison of reconstruction with PLIC and with the MIR model. 
 Reconstructions are created from the volume fraction image.}
 \label{fig:mir_examples}
\end{subfigure}
\begin{subfigure}{0.3\textwidth}
\centering
 \includegraphics[width=\columnwidth]{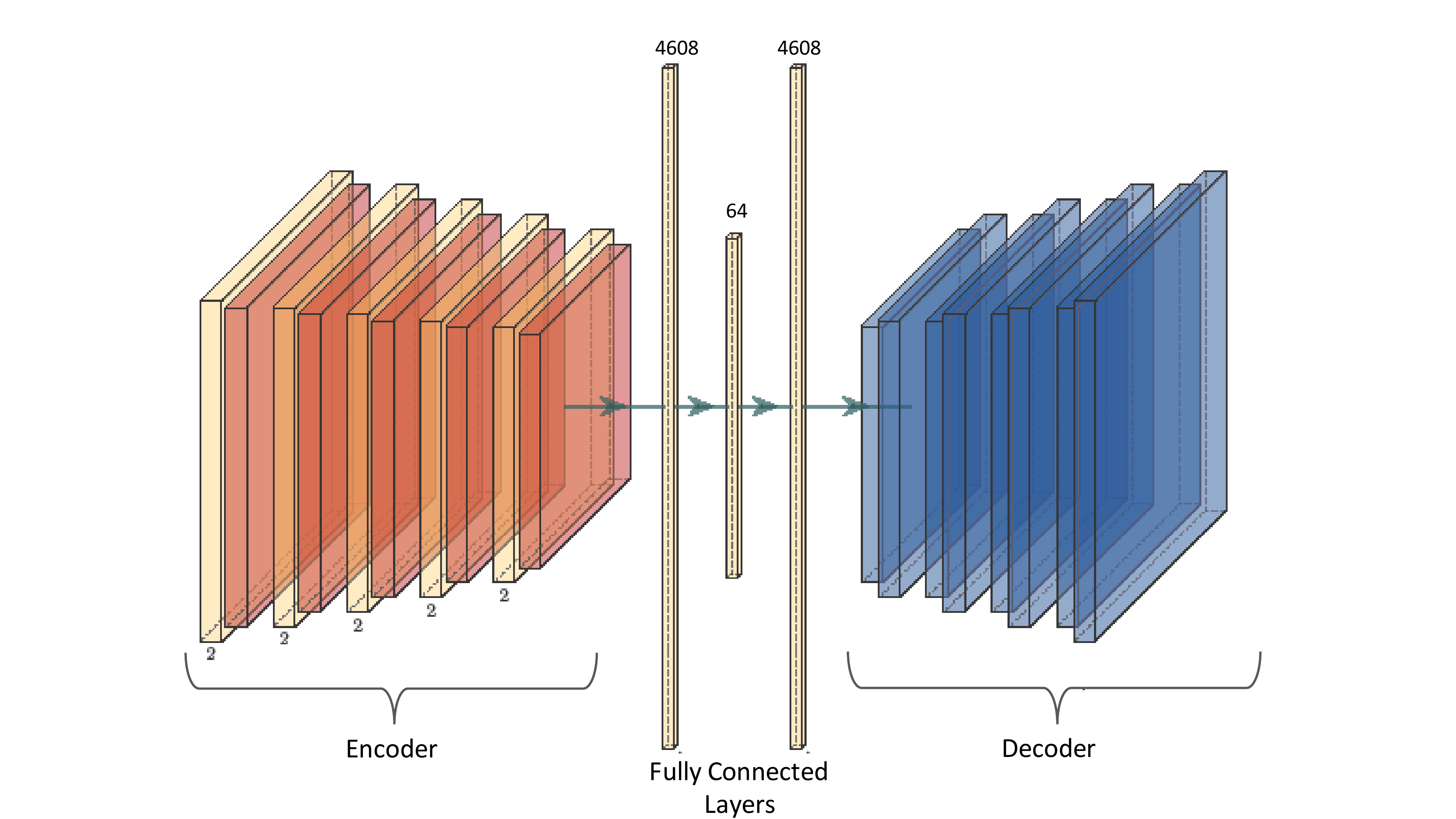}
 \caption{Diagram of MIR Model Architecture}
 \label{fig:MIR_model}
\end{subfigure}
\begin{subfigure}{0.3\textwidth}
\centering
 \includegraphics[width=\columnwidth]{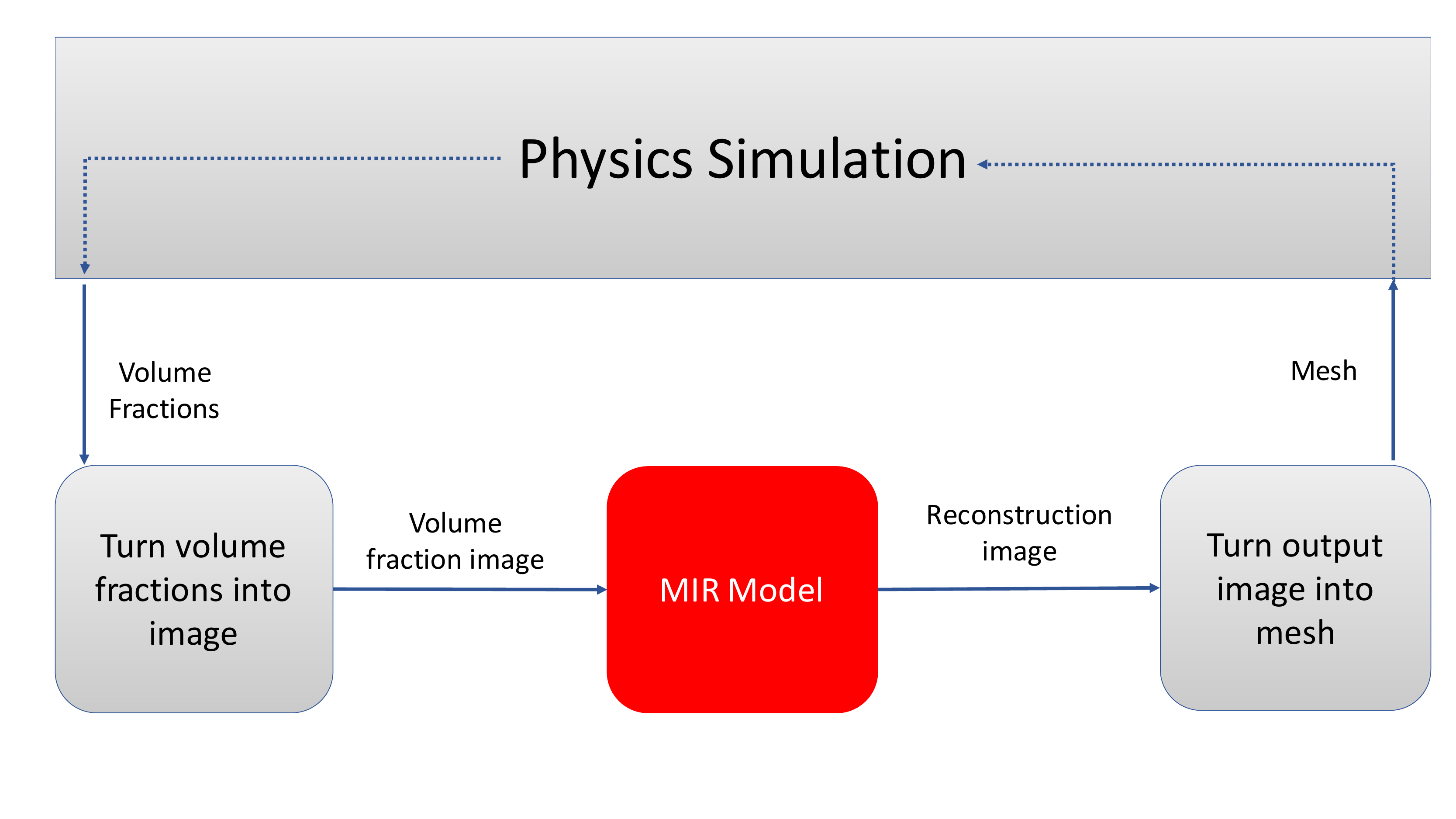}
 \caption{MIR model in physics simulation workflow}
 \label{fig:mir_workflow}
\end{subfigure}
\caption{Material Interface Reconstruction (MIR) Model}
\end{figure*}

The MIR model is a neural network used for material interface reconstruction in
physics simulations, in which boundaries are constructed between
immiscible materials based on volume fractions for each zone in the
environment. Current methods tradeoff continuity and conservation of material,
often to the detriment of the reconstruction accuracy. Figure
\ref{fig:mir_examples} compares outputs for one current reconstruction method,
PLIC, and the MIR model. PLIC conserves the volume of the material in each zone
but leads to discontinuous boundaries. The MIR model is able to create
continuous boundaries while conserving volumes and looks remarkably like the
ground truth.

The MIR model is a convolutional autoencoder and is
diagrammed in Figure~\ref{fig:MIR_model}. It is composed of
4 convolution layers with pooling, layernorm
after every convolution, 3 fully connected layers, two of which with 4608
neurons each, and transposed convolution layers to return the output of the fully
connected layers to the size of the original image. The weights of the convolution
and transposed convolution layers are tied as a form of regularization.
In total, there are 700K parameters in the MIR model.

Mixed zones, or zones with more than one material in them,
are processed each simulation timestep.
The workflow of the simulations is shown in Figure~\ref{fig:mir_workflow}.
The number of mixed zones per GPU for each simulation
timestep ranges from the thousands to the hundreds of thousands depending on
the application. The number of zones per timestep may vary throughout the
simulation in some applications. To not impede the physics calculations going
on around it, the target throughput of the model is 100,000 samples per second per MPI rank.

\subsection{Optimizing the MIR Model for dataflow architectures}

The original MIR model, which was much larger, did not translate well to a
dataflow architecture.
In order to optimize for dataflow, batchnorm layers
were replaced with layernorm layers and work was done to shrink particularly
large fully connected layers. The GPU model was subsequently changed to match
the version run on the DataScale system.

\section{Evaluation}

To test the efficacy of these new accelerators for \texttt{in-the-loop} inference
workloads we use the Hermit and MIR models, described in
Section~\ref{sec:models}. We measure the inference performance of 3 accelerator
architectures: 1) Nvidia GPUs, 2) AMD GPUs, and 3) SambaNova DataScale. We
describe our experimental setup, demonstrate optimizing performance on GPU and
dataflow architectures, and compare the architectures for \texttt{in-the-loop} CogSim
workloads.

\subsection{Experimental Setup}
\label{sec:experimental_setup}

In this section, we describe the experimental setup for each of the tested
architectures. We established a fair comparison across different architectures
by working with the vendors to ensure optimal configurations and accurate
measurements of the hardware performance. We implemented the models in PyTorch
and used them as references to generate the models on all tested hardware. For
the GPU tests, the models were run with the PyTorch API as well as used to
generate compiled models for Nvidia's TensorRT and CUDA Graphs APIs. On the
DataScale, the SambaFlow software stack compiles a model from the PyTorch
reference model.

In our experiments for all hardware, we measured half-precision inference
latency and throughput bandwidth. We performed these measurements across a
range of mini-batch sizes (i.e., 1, 4, 16, 64, 256, 1K, 2K, 4K, 8K, 16K, and
32K) to capture the performance landscape and record how it is modulated by the
number of samples being pushed through a model. We adjusted the total number of
samples run through a model, with respect to the mini-batch size, such that the
total wall-clock time is greater than 10s for each performance run. Latency was
measured in milliseconds (ms) as the average time across all mini-batches for
running inference on a single mini-batch of samples. Throughput was measured in
samples per second across all samples of a given mini-batch size. Before each
performance measurement on the GPUs, we ``warmed-up'' the hardware by running
inference on 10 mini-batches. The performance measurement occurred directly
after the warm-up phase.

We performed two types of experiments: node-local inference and remote
inference. As discussed in Section~\ref{sec:background}, surrogate models can
reside on the same GPU as the CogSim simulation. Therefore we measured
the node-local inference for GPUs, where the input data is generated and model
inference is executed on the same GPU. Also discussed in
Section~\ref{sec:background}, new dataflow architecture accelerators, such as
the SambaNova DataScale, can reside on separate nodes and be made available to
compute nodes via a high-speed InfiniBand network. We first measure node-local
inference on the DataScale to optimize model performance, sans network latency,
to obtain an upper bound on performance. We then measured remote inference
with the DataScale, where input data is generated on a compute node, sent
across the network to the DataScale node, and inference results are sent back
across the network to the originating compute node.  Communication was done via
a prototype C++ API and library, but these tests mimic the expected use case
where multiple MPI ranks would issue queries to the DataScale node.
Additionally, the GPU measurements include no data movement between system
memory and the accelerator to mimic the real-world CogSim application, where
the simulation and models are both resident on the GPU. Measurements on the
DataScale do include movement to and from system memory for both node-local and
remote inference.

We tested several generations of GPU architecture from Nvidia (i.e., P100,
V100, and A100) and AMD (i.e., MI50 and MI100). For each of these hardware
types, we use PyTorch 1.9.0. Our Nvidia setup uses CUDA 11.4, cuDNN 8.2.2.26,
and TensorRT 8.0.1.6. Our AMD setup uses ROCm 4.2 and MIOpen 2.11.0.  Models
were run in half-precision with FP16 on the GPUs. In addition to our baseline
PyTorch implementation, we also worked with the vendor to optimize performance
for an A100 GPU. We tested TensorRT and CUDA Graphs APIs in Python and C++.
TensorRT, with the Torch2TRT library, compiles our PyTorch model to optimize
performance through kernel selection and by combining layers. CUDA Graphs
allows the model to be called from a single CPU operation and reduces kernel
launch overhead from the PyTorch API. While TensorRT, CUDA Graphs, and C++ add
complexity to running the model, the benefit-cost ratio is high for a
frequently used inference model (e.g., a CogSim surrogate model).

For the dataflow architecture, we tested the DataScale AI accelerator from
SambaNova Systems.  Our surrogate models are run in half-precision with BF16 on
the DataScale. The SambaFlow 1.8 library generated a compiled model from the
PyTorch model. Given the relative novelty of this hardware, we performed a more
extensive survey of the performance landscape compared to GPU-related
experiments. The DataScale system is composed of 8 SN10 RDUs. Each RDU contains
4 tiles, which are discrete compute and memory units of the RDU.  A single
model can be deployed in various configurations, ranging from ${}^{1}/{}_{4}$
of an RDU (i.e., 1 tile) up to a complete RDU (i.e., 4 tiles).  The RDU uses an
additional \texttt{micro-batch size} parameter that the GPUs do not have. This
parameter determines the size of accumulated data sent across the RDU tiles
during inference. The micro-batch size must always be equal to or less than the
mini-batch size. In our exploration of DataScale performance, we tested how
inference performance scales with RDU tiles using different combinations of
mini-batch and micro-batch sizes. From these results we established a baseline
for PyTorch node-local performance of the DataScale and then worked with the
vendor to optimize the model using C++ APIs and hand-tuned model placement on
the hardware.

Remote inference performance tests on the DataScale were run with the
configurations (i.e., RDU tile count, mini-batch size, and micro-batch size,
optimizations) that provided the best performance. In remote inference
experiments, data is generated on a remote compute node, sent across a
high-speed network, inference is performed on the DataScale, and the results
are returned to the remote client. The high-speed network in our tests was the
Mellanox Infiniband ConnectX-6 with a bandwidth of 100 Gb/s. Latency
measurements include the additional round-trip of data transfer. Throughput was
maximized in these tests by allowing asynchronous communication between the
client and SN10-8 server.  The client sends mini-batch $n+1$ to the server
before inference results for mini-batch $n$ are returned to the client.

All experiment measurements were replicated 5 times. The figures in the
remainder of this section plot the mean of the 5 measurements with error bars
indicating the $95\%$ confidence interval.

\subsection{Tuning for the GPU architectures}
\label{sec:eval_gpu}

\newcommand{\NvidiaLatencyOneSample}{$0.65$}
\newcommand{\NvidiaLatencyMaxSample}{$3.92$}
\newcommand{\NvidiaThroughputOneSample}{$1{,}534$}
\newcommand{\NvidiaThroughputMaxSample}{$8.35M$}

\newcommand{\AMDLatencyOneSample}{$0.96$}
\newcommand{\AMDLatencyMaxSample}{$5.59$}
\newcommand{\AMDThroughputOneSample}{$1,043$}
\newcommand{\AMDThroughputMaxSample}{$5.85M$}

\newcommand{\TRTGraphsLatencyOneSample}{$0.12$}
\newcommand{\TRTGraphsLatencyMaxSample}{$1.52$}
\newcommand{\TRTGraphsThroughputOneSample}{$8{,}240$}
\newcommand{\TRTGraphsThroughputMaxSample}{$21.6M$}

\begin{figure}[!htb]
\centering
 \includegraphics[width=\columnwidth]{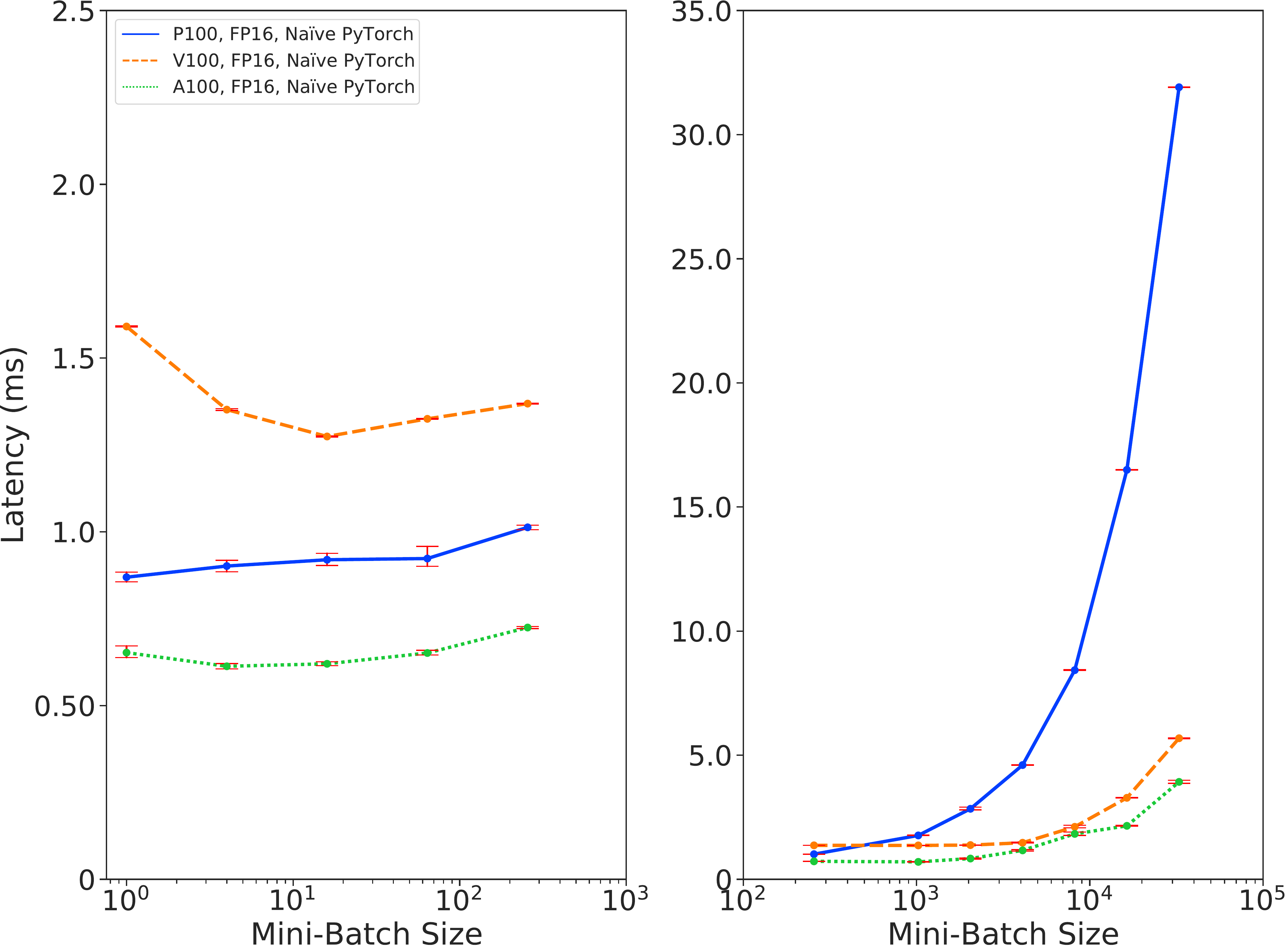}
 \caption{Inference latency of the Hermit model on Nvidia P100, V100, and
A100 GPUs using the PyTorch Python API. We observe the lowest latency across
all mini-batch sizes with the A100.}
 \label{fig:nvidia_scaling_latency}
\end{figure}

We measured performance of 2 surrogate models, Hermit and MIR, across 3
generations of Nvidia GPUs: P100, V100, and A100.
Figure~\ref{fig:nvidia_scaling_latency} shows the inference latency of the
Nvidia GPUs across different mini-batch sizes for the Hermit model. The left
panel of the figure shows a nearly constant latency for each of the GPUs at
mini-batch sizes smaller than 256. The A100 has the lowest single sample
latency of {\NvidiaLatencyOneSample}ms.  Perhaps unexpectedly, the V100 latency
is larger than the P100 at these small mini-batch sizes. The Hermit model is a
small network, so with the ``na\"{i}ve PyTorch'' implementation, performance at
small mini-batch sizes is CPU-bound.  This is due to the overhead of PyTorch
CPU logic that determines which GPU kernels to launch. In context of the
measured GPU performance, both the P100 and A100 systems use x86 architecture,
while the V100 system uses Power9. The CPU-bound nature of the na\"{i}ve
PyTorch implementation at small mini-batch sizes is the cause of larger latency
with the V100 compared to the P100 in Figure~\ref{fig:nvidia_scaling_latency}.

On the right side of Figure~\ref{fig:nvidia_scaling_latency}, we see that the
latency increases more rapidly for the P100 than either the V100 or A100. This
result suggests that at these larger mini-batch sizes for the Hermit model the
P100 hardware becomes saturated. The P100 latency is more than 8x that of the
A100 at the largest mini-batch size of 32K. The A100 has a latency of
{\NvidiaLatencyMaxSample}ms at this mini-batch size.

\begin{figure}[!htb]
\centering
 \includegraphics[width=\columnwidth]{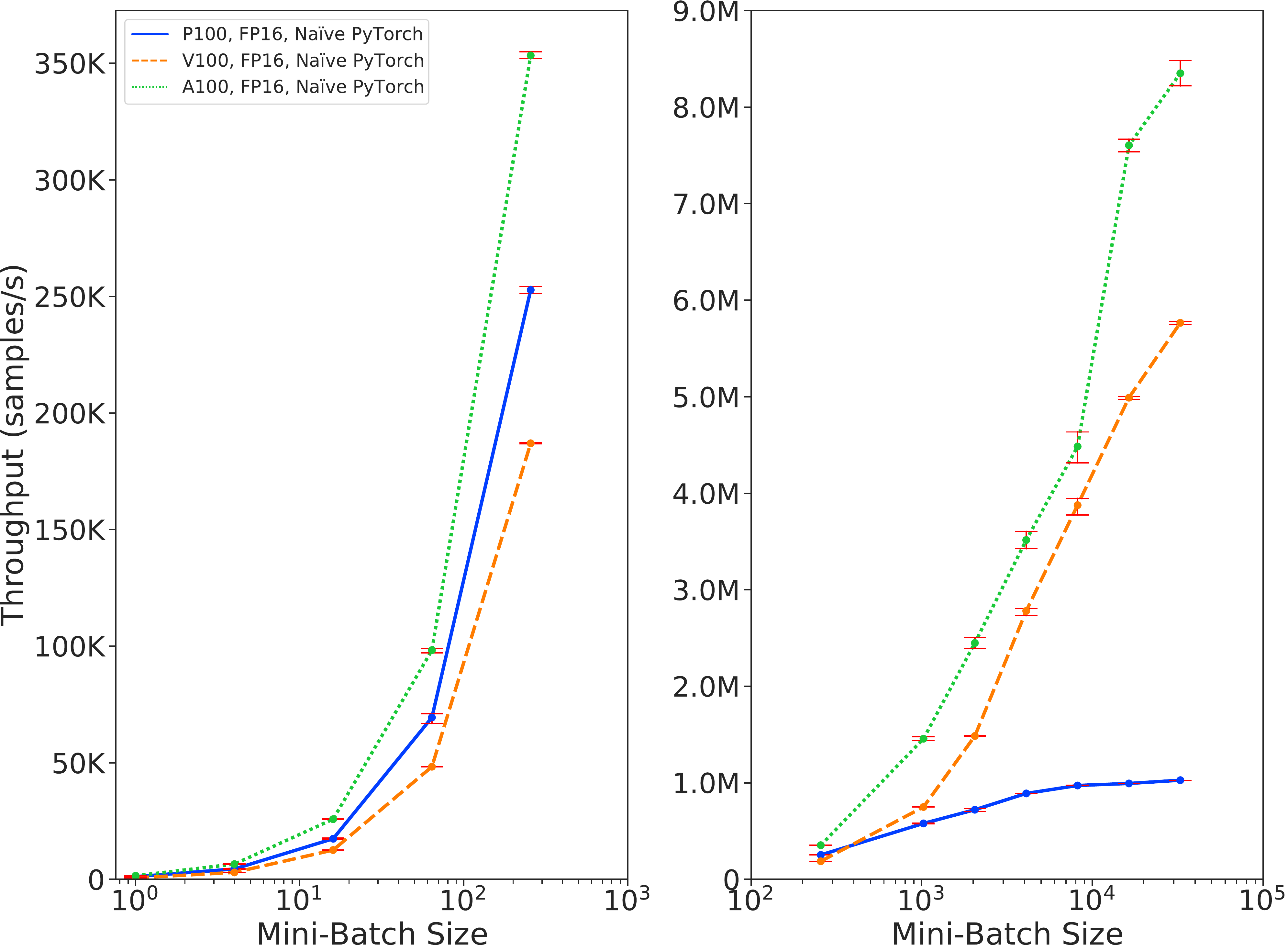}
 \caption{Inference throughput of the Hermit model on Nvidia P100, V100, and
A100 GPUs using the PyTorch Python API. We observe more than 8x speedup on the
A100 compared to the P100 at the largest batch sizes.}
 \label{fig:nvidia_scaling_throughput}
\end{figure}

Measured throughput of the three Nvidia GPUs with the Hermit model is shown in
Figure~\ref{fig:nvidia_scaling_throughput}. We see a similar pattern as the
latency measurements, with the V100 being slower than the P100 at the
smallest mini-batch sizes. At larger sizes the additional transistor and memory
hardware on the V100 and A100 are apparent as they achieve inference
throughputs in excess of 5 Million samples/s. The throughput measurements of
the A100 were largest for all mini-batch sizes, with 1 and 32k mini-batch
throughputs of {\NvidiaThroughputOneSample} and {\NvidiaThroughputMaxSample}
samples/s, respectively.

\begin{figure}[!htb]
\centering
 \includegraphics[width=\columnwidth]{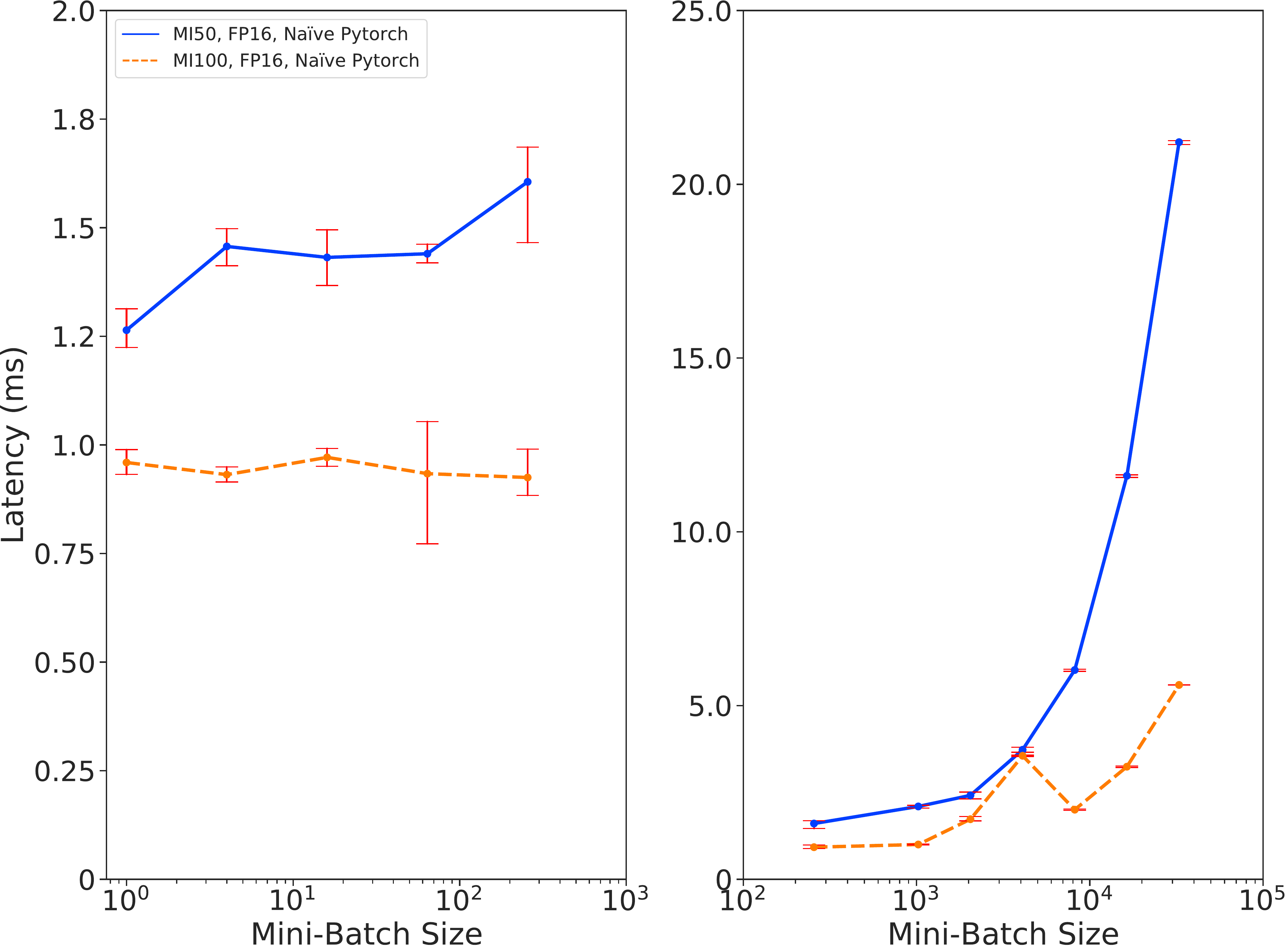}
 \caption{Inference latency of the Hermit model on AMD MI50 and MI100
GPUs using the PyTorch Python API. We observe the lowest latency across all
mini-batch sizes with the MI100.}
 \label{fig:amd_scaling_latency}
\end{figure}

In Figure~\ref{fig:amd_scaling_latency} we show the measured latency for Hermit
on two AMD GPUs, the MI50 and MI100. We observe near constant latency with the
MI100 for mini-batch sizes at and below 1K. Single sample latency of the MI100
is measured at {\AMDLatencyOneSample}ms. MI50 performance was similar to P100
performance in Figure~\ref{fig:nvidia_scaling_latency} as we see a marked
increase in latency as the mini-batch size increases beyond 1K. Our
measurements also show an unexpected drop in performance for the MI100 at a
mini-batch size of 4K. At the maximum mini-batch size of 32k, we recorded a
latency of {\AMDLatencyMaxSample}ms on the MI100, corresponding to a maximum
throughput of {\AMDThroughputMaxSample} samples/s.

\begin{figure}[!htb]
\centering
 \includegraphics[width=\columnwidth]{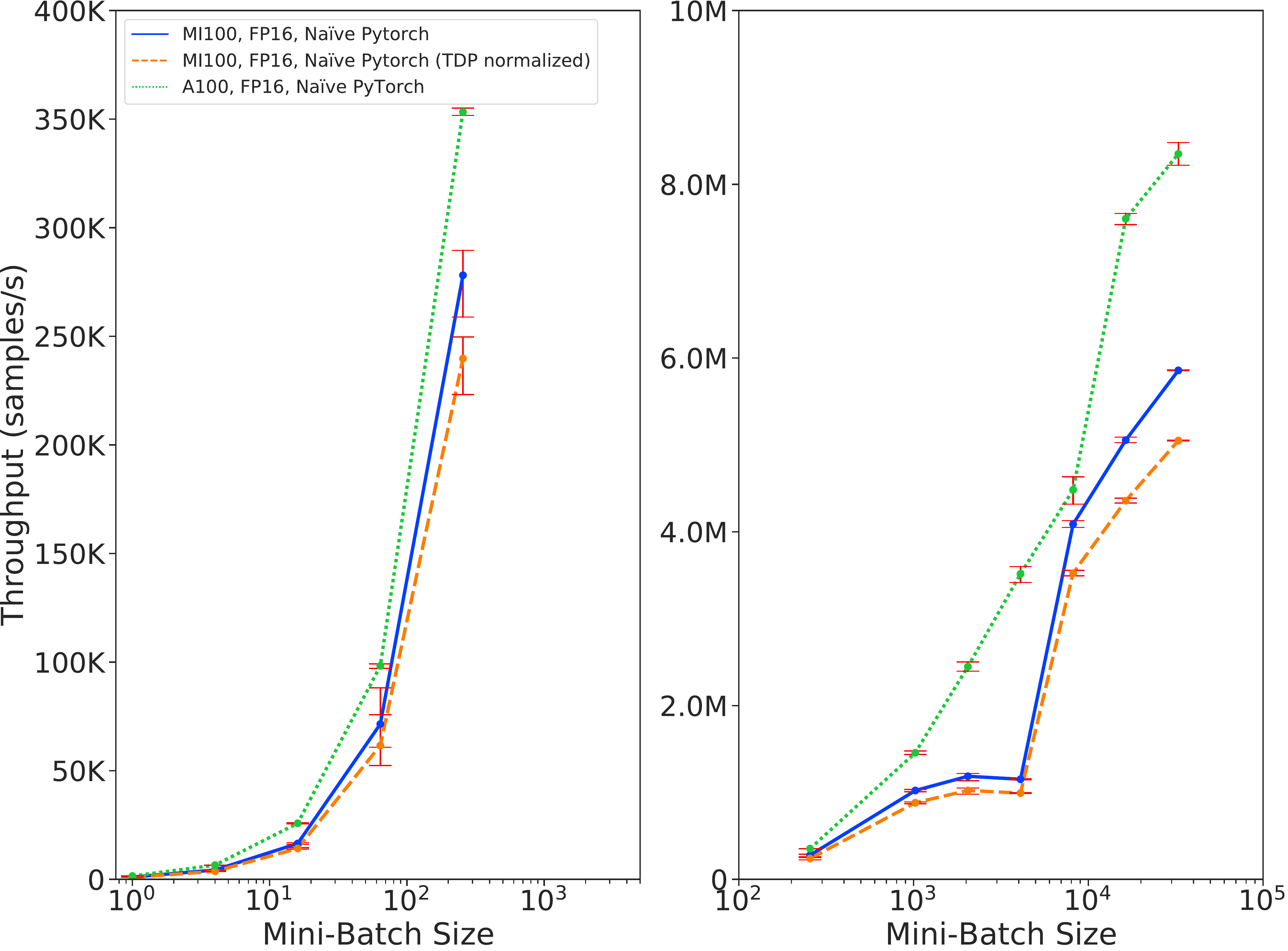}
 \caption{Inference latency of the Hermit model on the Nvidia A100 and
AMD MI100.}
 \label{fig:gpu_compare_throughput}
\end{figure}

A comparison of the fastest Nvidia GPU (i.e., A100) and fastest AMD GPU (i.e.,
MI100) is shown in Figure~\ref{fig:gpu_compare_throughput}. We observe that the
measured throughput of the A100 is larger than the MI100 at all tested
mini-batch sizes.  At the 32K mini-batch size, the A100 can process more than
2M additional samples per second than the MI100. We also note that the A100 has
a lower TDP at 250W than the MI100 at 290W. We normalize throughput of the
MI100 based on TDP and show these values in
Figure~\ref{fig:gpu_compare_throughput} as well. At the smallest mini-batch
sizes the A100 is superior to the MI100 with respect to latency, with measured
single sample latencies of {\NvidiaLatencyOneSample}ms and
{\AMDLatencyOneSample}ms.  The unexpected plateau in performance of the MI100
between mini-batch sizes of 1K and 4K shown in
Figure~\ref{fig:gpu_compare_throughput} may be explained by the beta support
for AMD GPUs of PyTorch 1.9.0.

\begin{figure}[!htb]
\centering
 \includegraphics[width=\columnwidth]{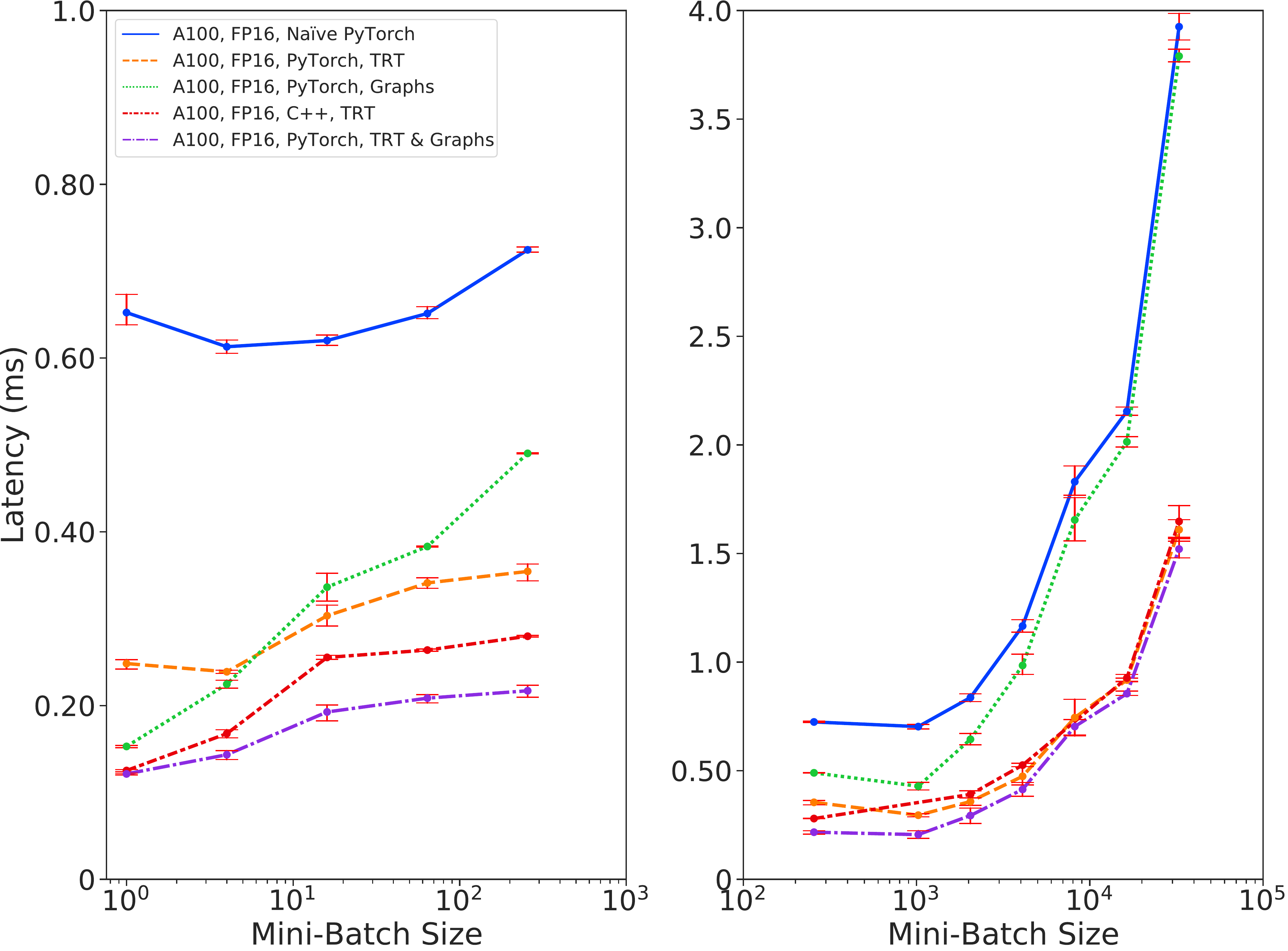}
 \caption{Inference latency of the Hermit model on a Nvidia A100 GPU using
combinations of Python and C++ APIs for PyTorch, CUDA Graphs, and TensorRT.}
 \label{fig:nvidia_compare_latency}
\end{figure}

Given the baseline performance measurements that we obtained, we next optimized
the performance of the Hermit and MIR models for the Nvidia A100 GPU. We worked
closely with the vendor to identify and implement optimizations that would
improve latency and throughput in our experimental measurements. We tested five
configurations: (1) na\"{i}ve PyTorch implementation, (2) PyTorch with TensorRT
using the torch2trt library, (3) PyTorch with CUDA Graphs, (4) PyTorch with
TensorRT and CUDA Graphs, and (5) C++ TensorRT API.

Figure~\ref{fig:nvidia_compare_latency} shows the measured inference latency
with these five configurations on an A100 GPU with the Hermit model. The left
panel of Figure~\ref{fig:nvidia_compare_latency} shows that all configurations
are more than twice as fast as the initial na\"{i}ve PyTorch implementation for
single sample latency. PyTorch with TensorRT and CUDA Graphs provides the
lowest inference latency for all mini-batch sizes, with a single sample latency
of {\TRTGraphsLatencyOneSample}ms and a 32k samples latency of
{\TRTGraphsLatencyMaxSample}ms.

\begin{figure}[!htb]
\centering
 \includegraphics[width=\columnwidth]{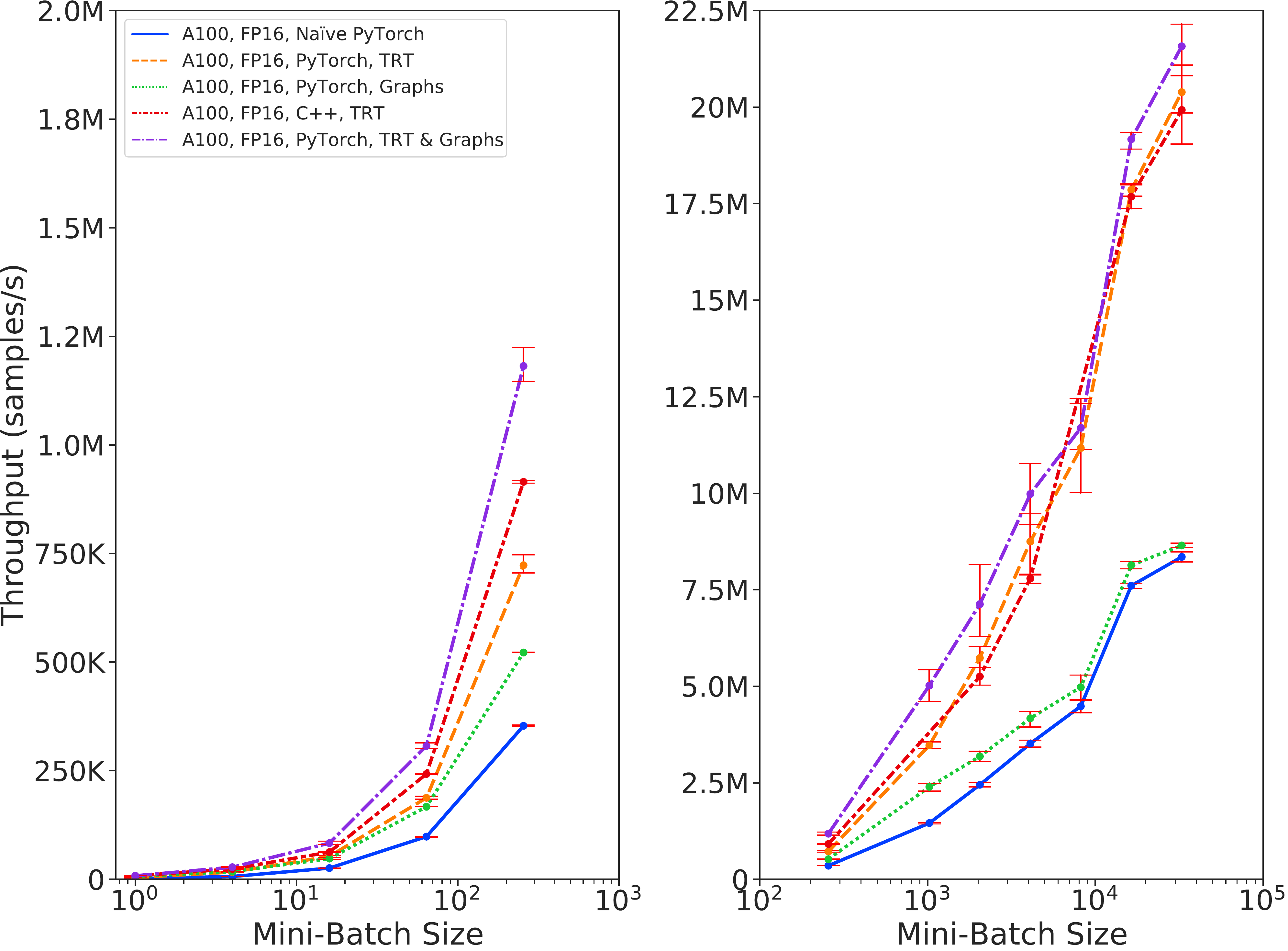}
 \caption{Inference throughput of the Hermit model on an Nvidia A100 GPU using
combinations of Python and C++ APIs for PyTorch, CUDA Graphs, and TensorRT.}
 \label{fig:nvidia_compare_throughput}
\end{figure}

\begin{figure}[!htb]
\centering
 \includegraphics[width=\columnwidth]{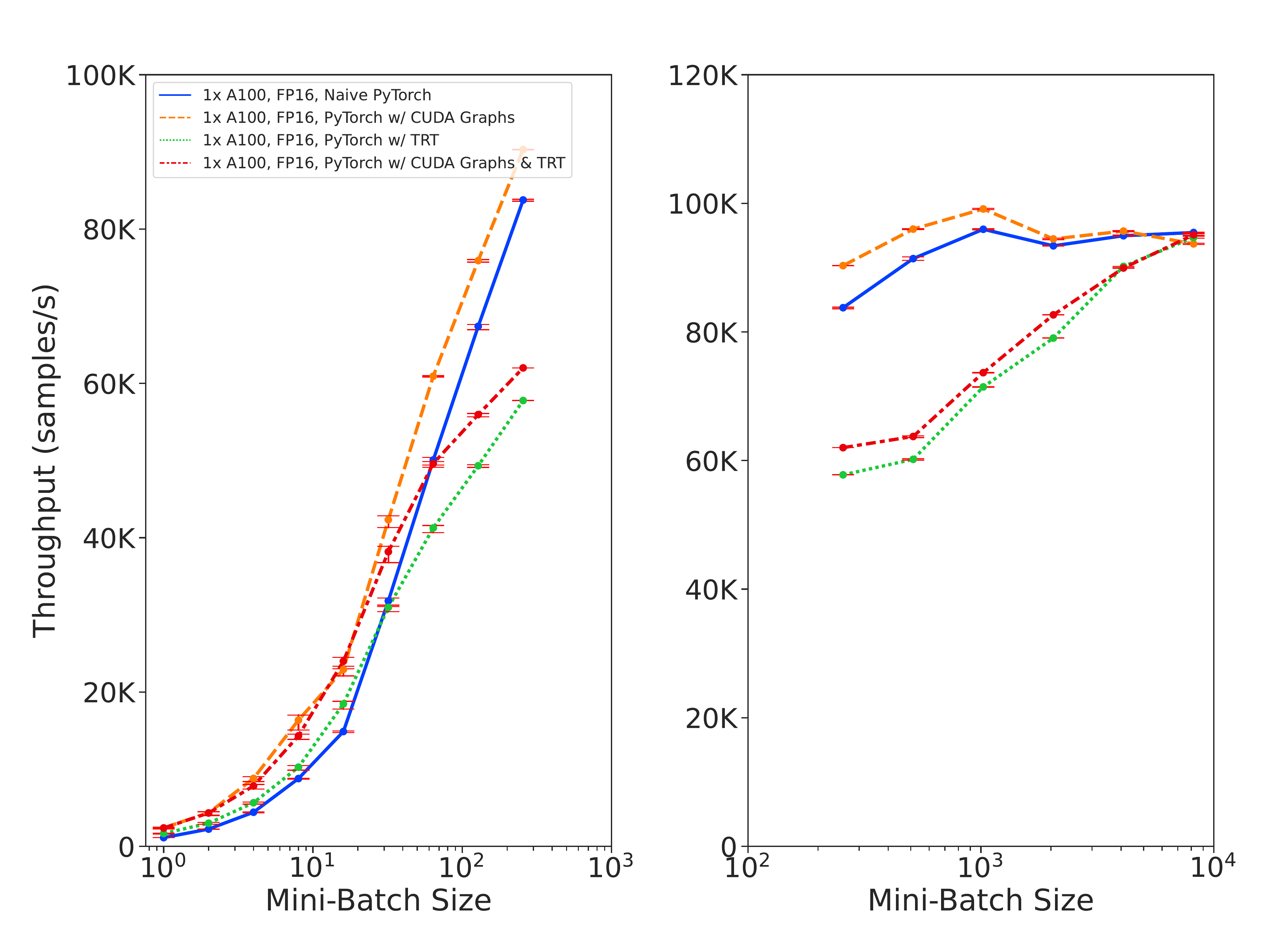}
 \caption{Throughput of the MIR Model using combinations of PyTorch, CUDA
 Graphs, and TensorRT APIs}
 \label{fig:mir_nvidia_optimizations}
\end{figure}

Throughput measurements for different configurations on the A100 with the
Hermit model are presented in Figure~\ref{fig:nvidia_compare_throughput}.
PyTorch with TensorRT and CUDA Graphs had the largest bandwidth for all
mini-batch sizes. We measured a single sample and 32K sample throughput of
{\TRTGraphsThroughputOneSample} samples/s and {\TRTGraphsThroughputMaxSample}
samples/s. We observe that all the configurations using TensorRT provide very
similar bandwidth performance across the tested mini-batch sizes.

We also measured performance of the MIR model on an A100 with 4 of the
described configurations.  Figure~\ref{fig:mir_nvidia_optimizations} shows that
for this larger model, CUDA Graphs gives the greatest increase in throughput
bandwidth.  This figure also shows that configurations using TRT have
measurably worse performance than the standard PyTorch implementation at
mini-batch sizes larger than 64. We note that this is caused by the Torch2TRT
library that we use to convert our PyTorch model to a TensorRT model. The
library has unoptimized implementations of \texttt{layernorm} and unary
functions that cause a performance bottleneck in the configurations using
TensorRT. The vendor has indicated that the upcoming TensorRT release will
address the performance issues we show here.
Figure~\ref{fig:mir_nvidia_optimizations} shows, that unlike with the Hermit
model, the MIR model performance on the A100 with different configurations
converge at the largest mini-batch size to nearly equal throughput bandwidth.
This may indicate that at the largest tested mini-batch size, the compute
capability of the A100 is saturated, regardless of the configuration.

\subsection{Tuning for a data flow architecture}
\label{sec:eval_sn}

\newcommand{\SNLatencyFourSample}{$0.04$}
\newcommand{\SNLatencyMaxSample}{$1.65$}
\newcommand{\SNThroughputMaxSample}{$8.14M$}

\newcommand{\SNLatencyRemoteFourSample}{$0.05$}
\newcommand{\SNThroughputRemoteMaxSample}{$6.4M$}

\begin{figure}[!htb]
\centering
 \includegraphics[width=\columnwidth]{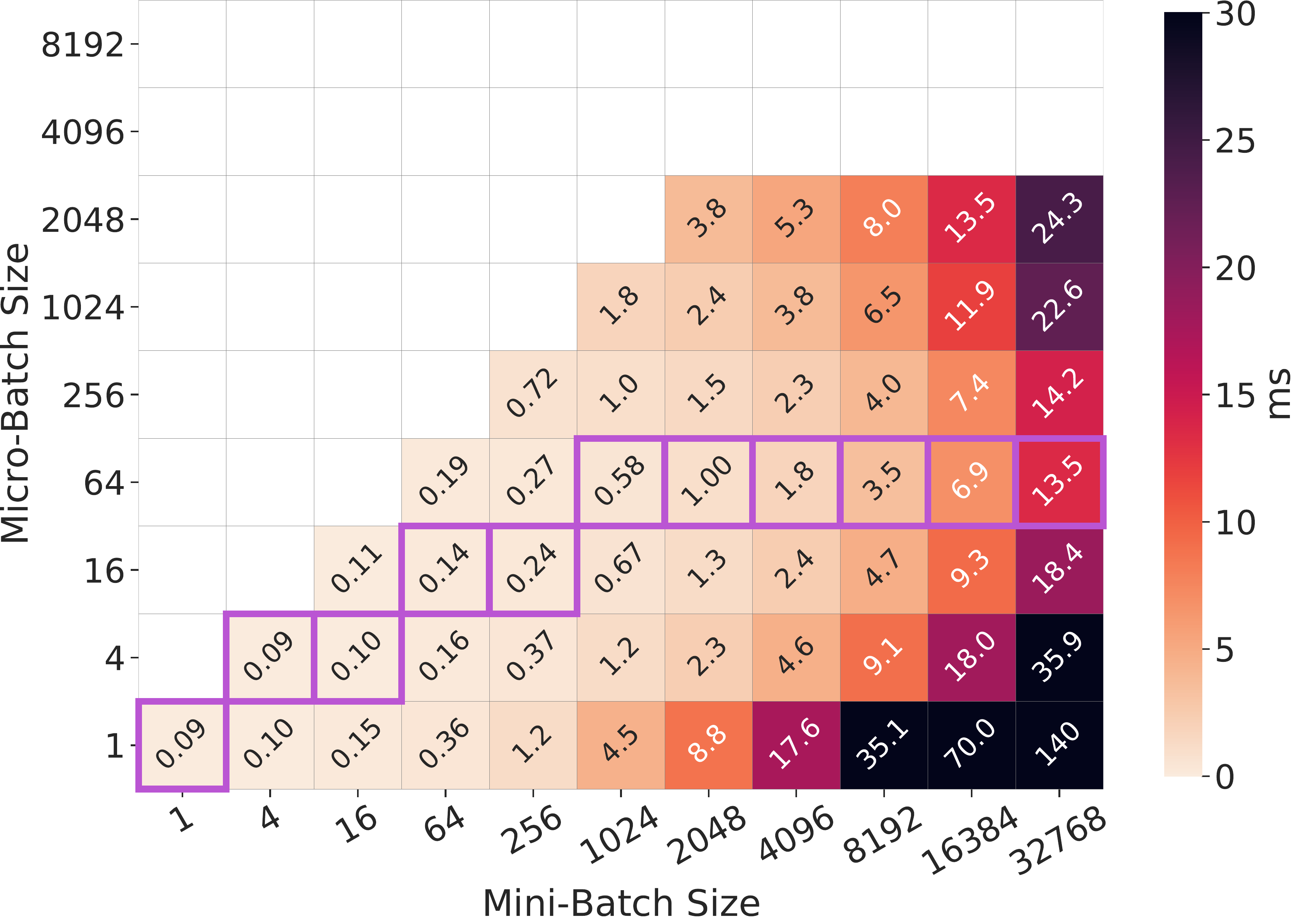}
 \caption{Latency of the Hermit model on ${}^{1}/{}_{4}$ RDU for
a range of mini-batch and micro-batch sizes using the Python API. The lowest
latency (mini-batch, micro-batch) pair for each mini-batch size is
highlighted.}
 \label{fig:sn_heatmap_latency_1}
\end{figure}

As discussed in Section~\ref{sec:experimental_setup}, the DataScale has several
tunable parameters. We measured the node-local performance of the Hermit model
on a different amount of RDU compute resources (i.e., RDU tiles) and across a
range of micro-batch and mini-batch sizes.
Figure~\ref{fig:sn_heatmap_latency_1} shows the latency of inference requests
with mini-batch sizes and micro-batch sizes ranging from 1 to 32K on a single
RDU tile (i.e., ${}^{1}/{}_{4}$ RDU).  In Figure~\ref{fig:sn_heatmap_latency_1}
we see that both mini-batch size and micro-batch size modulate the latency of
the network on the DataScale.  Each mini-batch size has a micro-batch size that
provides optimal performance in terms of latency. We highlight the minimum
latency values in purple for each mini-batch size. We also note that tuning the
micro-batch and mini-batch size to be multiples of 6 offers additional
performance in terms of latency and throughput by exploiting hardware
properties of the DataScale. We show this below in
Figures~\ref{fig:sn_compare_latency}~and~\ref{fig:sn_compare_throughput}.
However, for the purposes of these tests, we focus on common mini-batch sizes
that are also optimized for GPU architectures.

\begin{figure}[!htb]
\centering
 \includegraphics[width=\columnwidth]{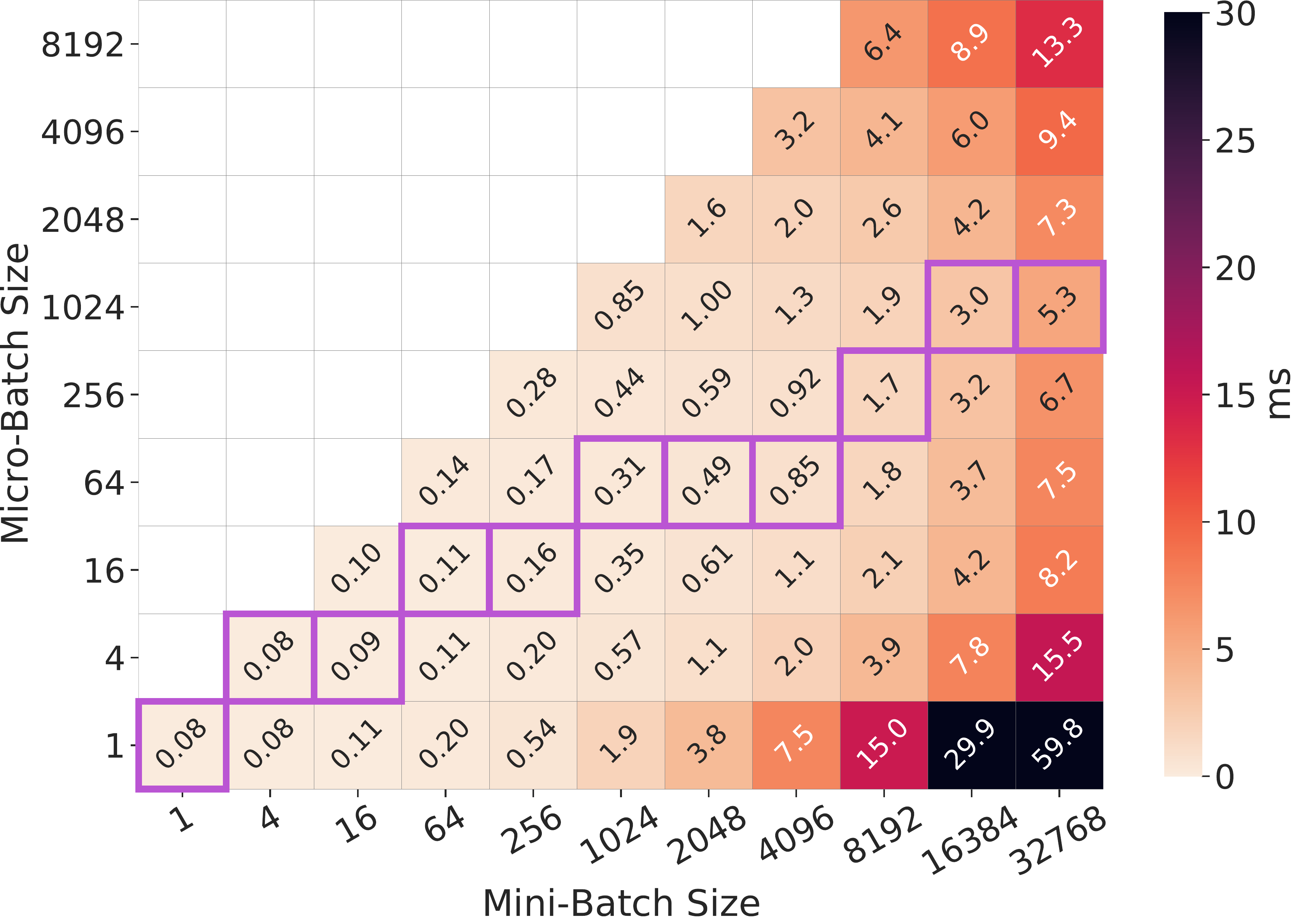}
 \caption{Latency of the Hermit model on 1 RDU for a range of
mini-batch and micro-batch sizes using the Python API. The lowest latency
(mini-batch, micro-batch) pair for each mini-batch size is highlighted.}
 \label{fig:sn_heatmap_latency_4}
\end{figure}

Figure~\ref{fig:sn_heatmap_latency_4} shows the latency of inference requests
with the same range of mini-batch size and micro-batch size as
Figure~\ref{fig:sn_heatmap_latency_1} but on 4 RDU tiles (i.e., one RDU).
Again, we highlight the combination of mini-batch size and micro-batch size
that give optimal performance. We observe that providing more RDU tiles for
model inference changes which mini-batch and micro-batch size combinations give
optimal performance.

In both Figure~\ref{fig:sn_heatmap_latency_1} and
Figure~\ref{fig:sn_heatmap_latency_4}, the white squares indicate
configurations that are not valid (e.g., micro-batch size larger than
mini-batch size) or failed to run on the hardware. From both figures, we can
see that at low mini-batch sizes, the micro-batch size has benign effects on
performance. As the mini-batch size increases, choosing the proper micro-batch
size is important for optimizing performance. For example, in
Figure~\ref{fig:sn_heatmap_latency_4} at a mini-batch size of 32K, the
difference between the slowest and fastest micro-batch size is 10-fold.  For
this reason, we performed parameter sweeps of the \texttt{(mini-batch,
micro-batch)} landscape for each tested configuration and report the maximum
throughput and minimum latency for each mini-batch size in the remainder of
this section.

\begin{figure}[!htb]
\centering
 \includegraphics[width=\columnwidth]{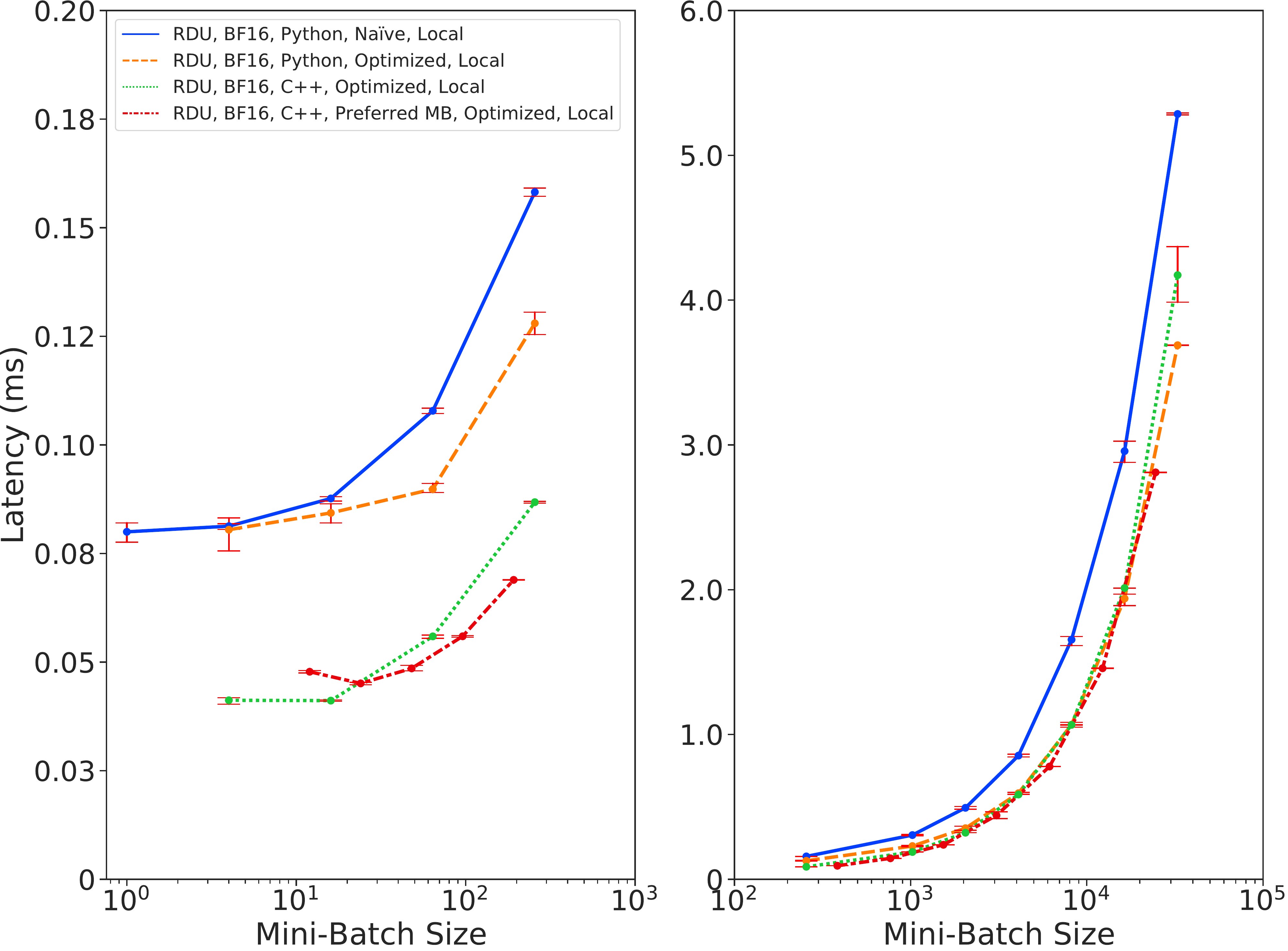}
 \caption{Inference latency of the Hermit model on the DataScale with 1 RDU
and 3 optimization methods.}
 \label{fig:sn_compare_latency}
\end{figure}

Similar to the Nvidia A100 GPU, the DataScale has several configuration options
that can optimize the model for latency and throughput.  We tested the Python
and C++ APIs and hand-optimized model placement for the DataScale to optimize
node-local inference.  Figure~\ref{fig:sn_compare_latency} shows the latency of
the Hermit model on 1 RDU with several configurations that optimize for
performance. We show the base performance of the Python API (i.e.,
``na\"{i}ve'') and show how different modifications improve overall
performance. Hand-optimized model placement (i.e., ``optimized'') on the
hardware provides benefits to the latency, especially at larger mini-batch
sizes. We also show that switching to a C++ API (with hand-optimized model
placement) provides additional benefits to latency. These benefits are
significant for the smallest mini-batch sizes, where inference latency is more
than halved compared to the Python API.

An additional optimization we show in Figure~\ref{fig:sn_compare_latency} is
from making small adjustements to the mini-batch and micro-batch sizes. In
Figure~\ref{fig:sn_heatmap_latency_4} we showed the performance landscape for
mini-batch and micro-batch sizes that are powers of 2 (i.e., $sizes \subset
2^{n}$). However, the DataScale hardware design is such that multiples of 6 for
micro-batch sizes and mini-batch sizes that are multiples of a given
micro-batch size can provide better performance. We show the effect of using
``preferred MB'' in conjunction with the other optimizations in
Figure~\ref{fig:sn_compare_latency}. The lowest latency values are observed
with the C++ API and hand-optimized model placement, with the exception of the
2 largest mini-batch sizes, where the Python API provides slightly lower
latency. At the smallest mini-batch sizes we observe a minimum latency of
{\SNLatencyFourSample}ms. The ``preferred MB'' optimzation provides additional
reduction in latency.

\begin{figure}[!htb]
\centering
 \includegraphics[width=\columnwidth]{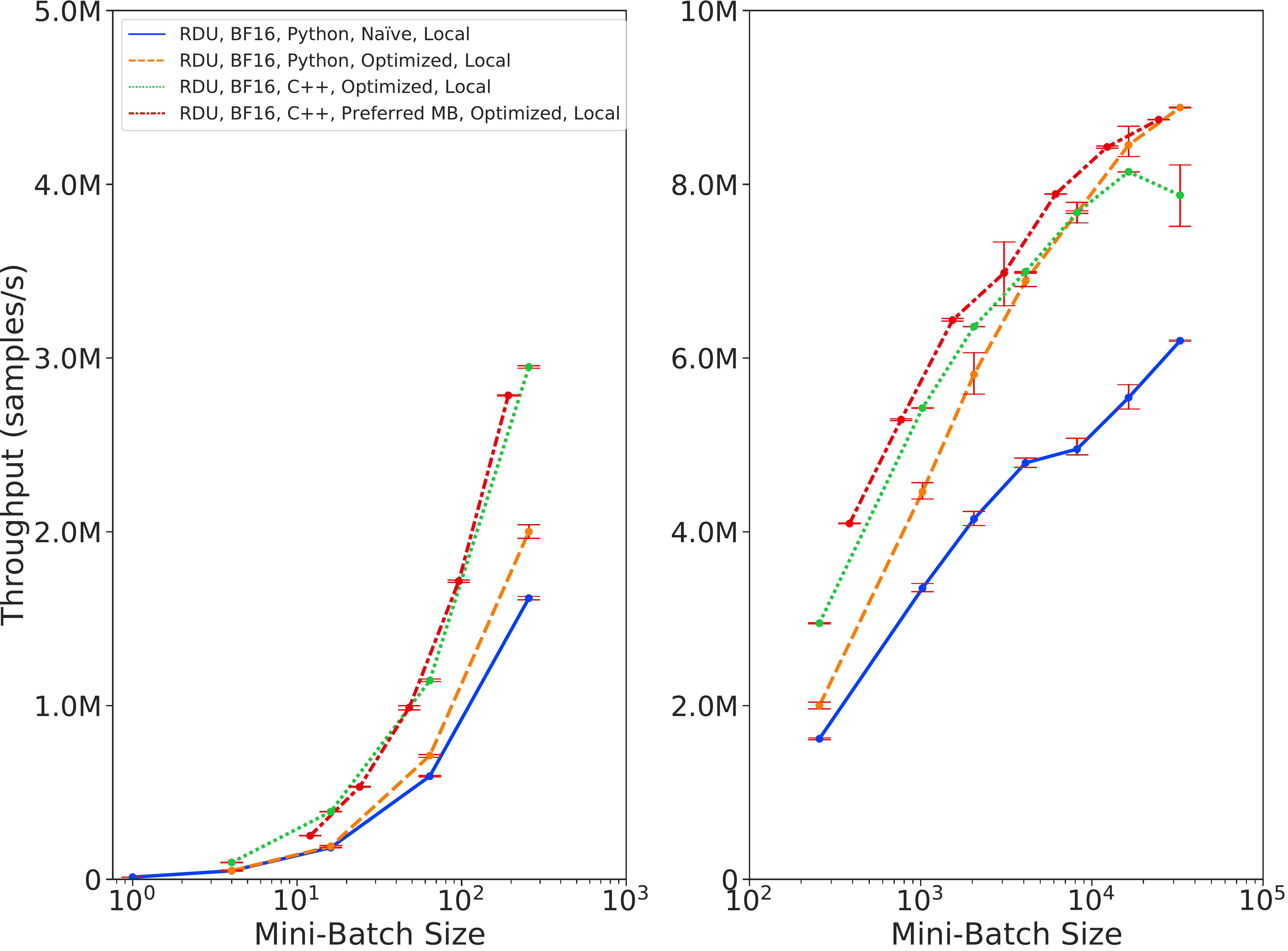}
 \caption{Inference throughput of the Hermit model on the DataScale with 1 RDU
and 3 optimization methods.}
 \label{fig:sn_compare_throughput}
\end{figure}

We also measured the throughput bandwidth of the different configuration of the
DataScale. Figure~\ref{fig:sn_compare_throughput} shows this data. On the right
panel of the figure, we see that the C++ API with hand-optimized model
placement provides a maximum throughput bandwidth of {\SNThroughputMaxSample}
samples/s at a mini-batch size of 16K. While the ``preferred MB'' optimzation
shows better latency and throughput, to maintain a fair comparison with GPU
architectures we show only the powers of 2 mini-batch sizes in subsequent
figures.  Figure~\ref{fig:sn_compare_latency} and
Figure~\ref{fig:sn_compare_throughput} show that the C++ API with
hand-optimized model placement provided the best overall performance on a
single RDU in node-local performance tests. We use this configuration for the
remote inference experiments. As such, the node-local performance will serve as
an upper limit for the performance we can expect from remote inference.

\begin{figure}[!htb]
\centering
 \includegraphics[width=\columnwidth]{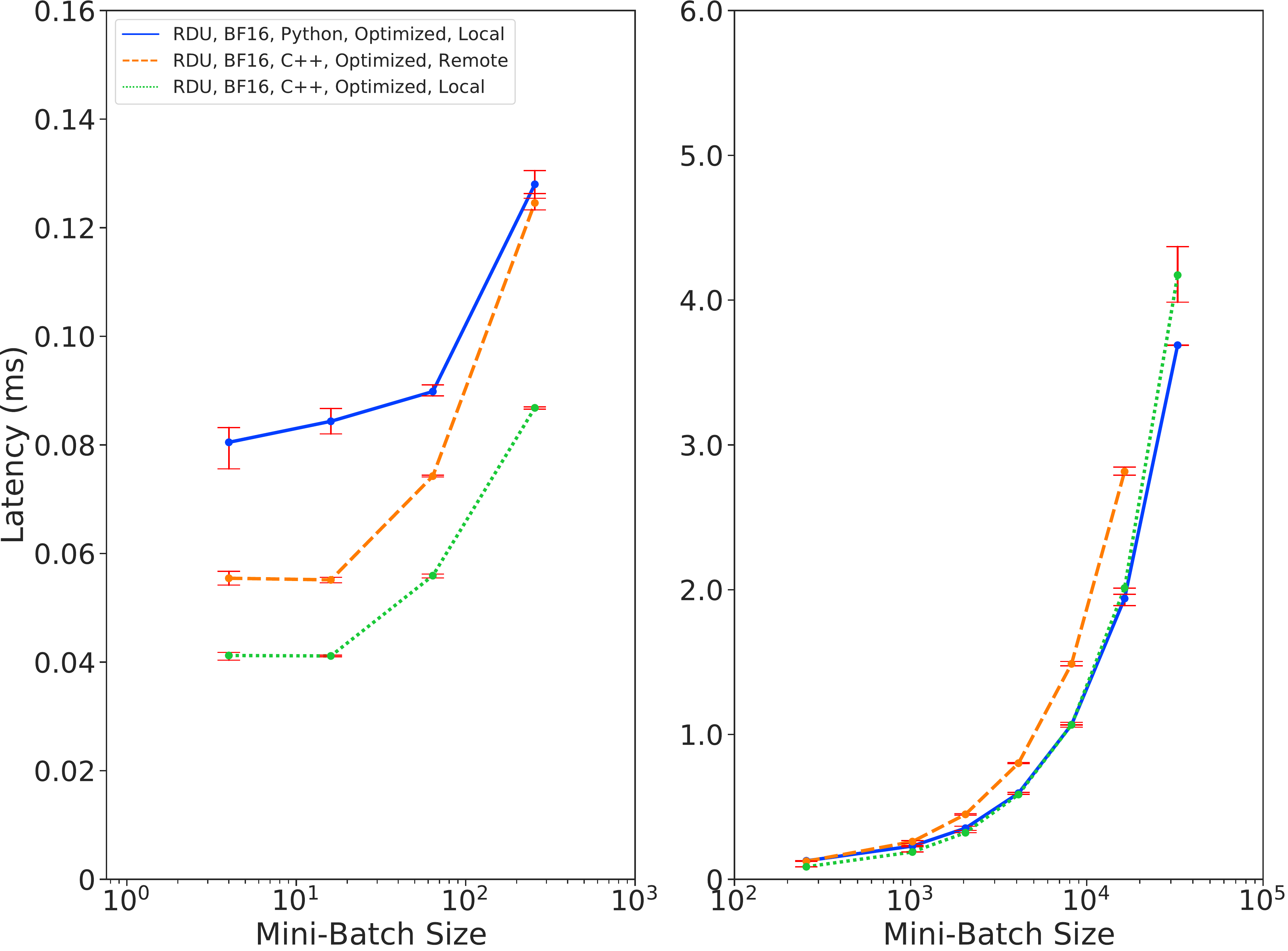}
 \caption{Inference latency of the Hermit model on the DataScale with 1 RDU
using the hand-optimized model placement and C+ API for local and remote
inference.}
 \label{fig:sn_remote_compare_latency}
\end{figure}

Because the DataScale System will be disaggregated from other system compute
nodes, the remote inference experiments provide information about expected
performance with CogSim applications that do \texttt{in-the-loop} inference.
Figure~\ref{fig:sn_remote_compare_latency} shows the remote inference latency
measurements for the Hermit model compared to node-local latency measurements.
All models were run with the hand-optimized model placement, the remote
inference is through the C++ API, and both Python and C++ APIs are shown for
node-local. In Figure~\ref{fig:sn_remote_compare_latency} we observe that
remote inference adds additional latency overhead compared to both Python and
C++ node-local inference. However, at the smallest batch sizes on the left
panel of Figure~\ref{fig:sn_remote_compare_latency} we see that C++ remote
inference can be as fast or faster than Python node-local inference, with an
average four sample latency of {\SNLatencyRemoteFourSample}ms. At a mini-batch
size of 16K, we observe the largest difference in performance between the
node-local and remote inference with the C++ API at $1.14$ms.

\begin{figure}[!htb]
\centering
 \includegraphics[width=\columnwidth]{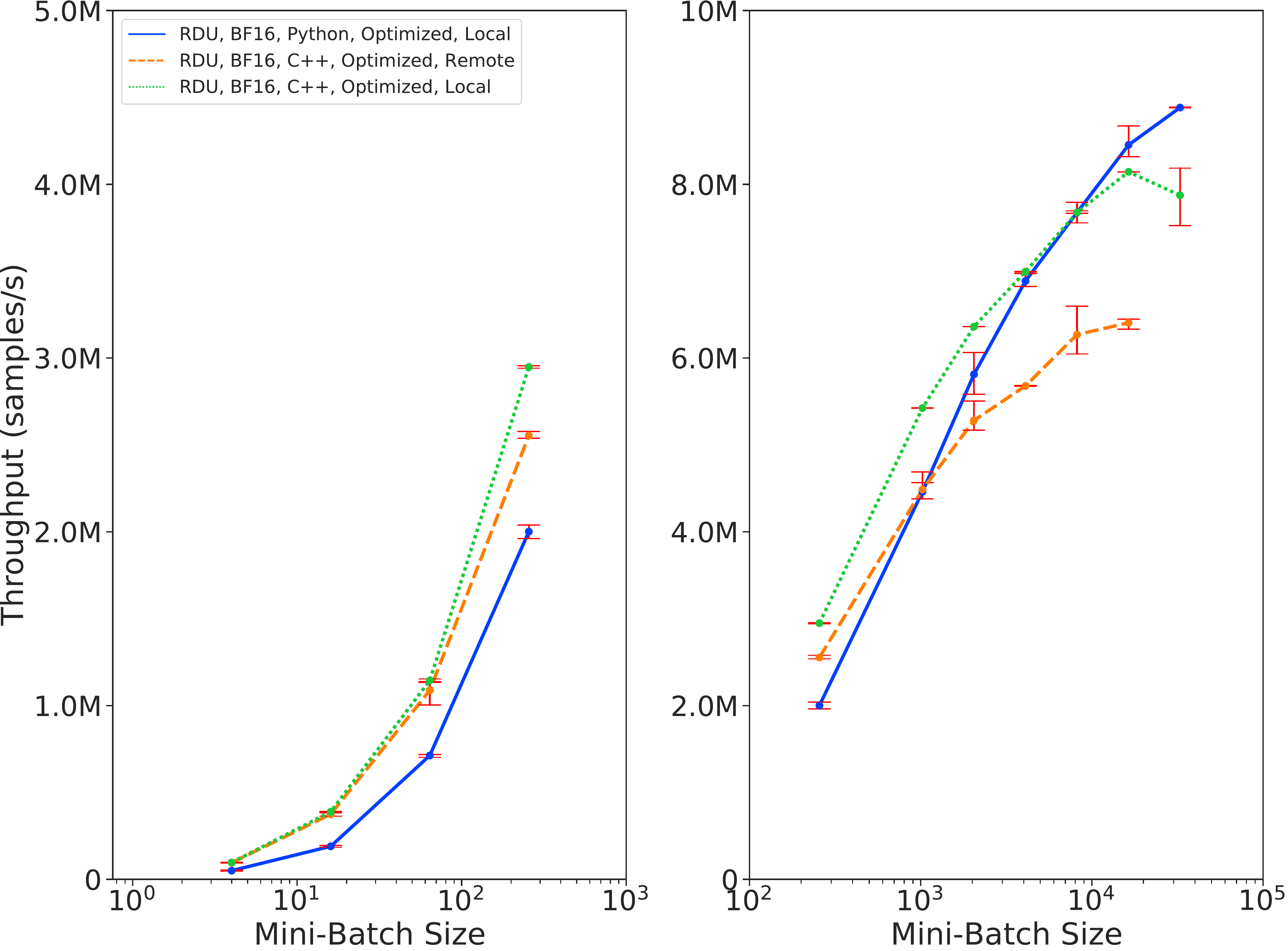}
 \caption{Inference throughput of the Hermit model on the DataScale with 1 RDU
using the hand-optimized model placement and C+ API for local and remote
inference.}
 \label{fig:sn_remote_compare_throughput}
\end{figure}

Remote inference throughput measurements for the DataScale are shown in
Figure~\ref{fig:sn_remote_compare_throughput}. In the left panel of the figure,
we see that the C++ remote inference throughput was between node-local Python
and node-local C++ inference measurements for small mini-batch. At mini-batch
sizes greater than 1K, both node-local configurations exceeded the remote
inference throughput. At a mini-batch size of 16K, a maximum remote inference
throughput of {\SNThroughputRemoteMaxSample} samples/s was recorded. Given the
competitive performance of the DataScale remote inference compared to
node-local performance shown in Figure~\ref{fig:sn_remote_compare_latency} and
Figure~\ref{fig:sn_remote_compare_throughput}, we find a compelling reason to
support disaggregated systems for CogSim workloads.  In the following sections,
we directly compare the performance of the dataflow architecture with the more
traditional GPU accelerators.

\subsection{Latency Comparison}
\label{sec:eval_latency}

The latency of inference is an important measurement for CogSim applications.
With \texttt{in-the-loop} inference, the time to return an inference result needs to be
short enough to avoid bottlenecking and degrading performance of the
application. In Section~\ref{sec:eval_gpu} we measured the performance of GPU
node-local inference in the context of a CogSim application running a surrogate
model. And in Section~\ref{sec:eval_sn} we measured the performance of the
DataScale node-local and remote inference. For disaggregated accelerators, the
remote inference is the measurement important to CogSim workflows.

\begin{figure}[!htb]
\centering
 \includegraphics[width=\columnwidth]{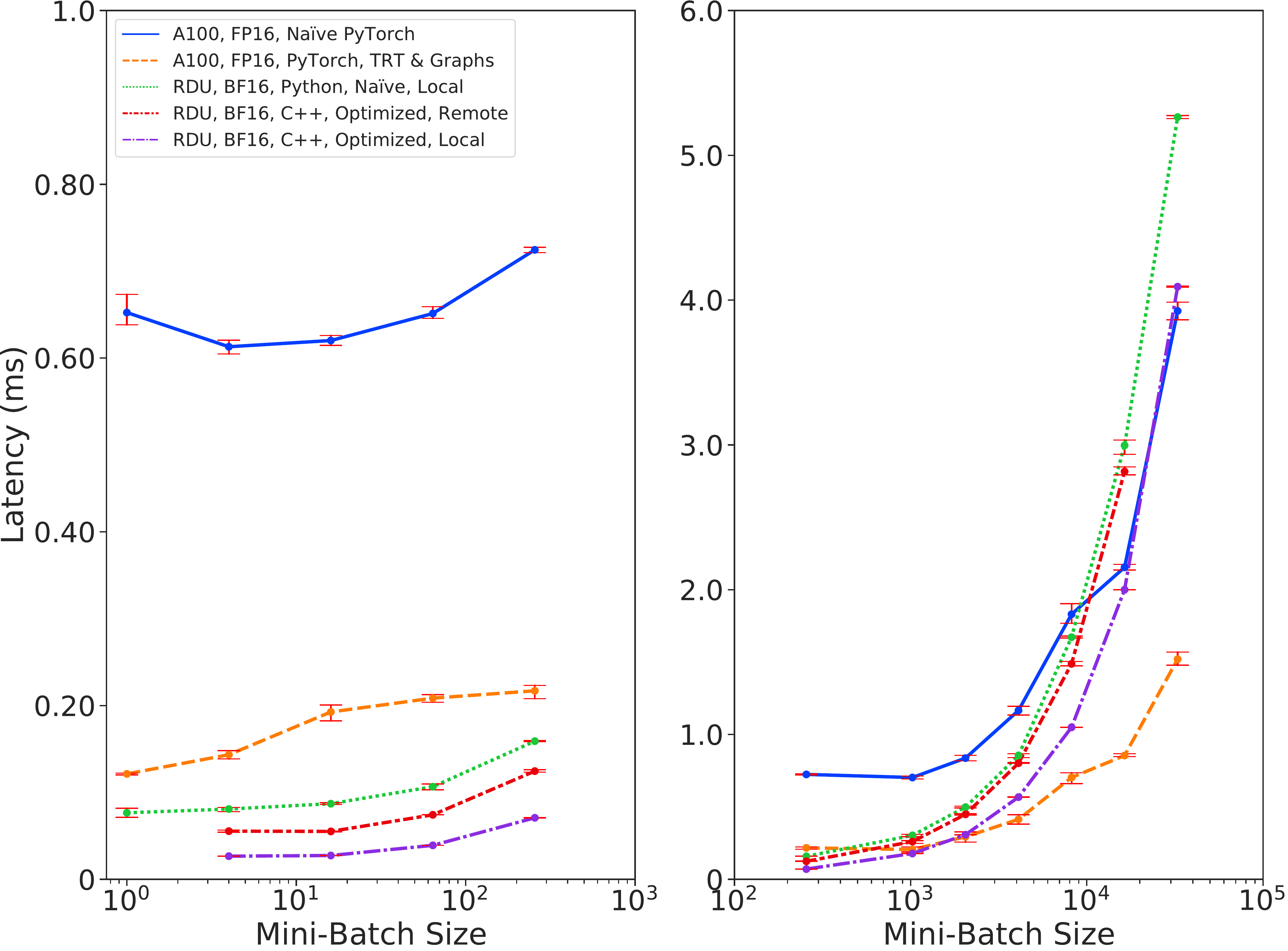}
 \caption{Inference latency of the Hermit model on the 1 RDU and A100 with
various configurations.}
 \label{fig:sn_gpu_compare_latency}
\end{figure}

Figure~\ref{fig:sn_gpu_compare_latency} shows a comparison of the mini-batch
latency measured for different configurations of the A100 and DataScale, from
Sections~\ref{sec:eval_gpu} and~\ref{sec:eval_sn}.  The included configurations
represent the fastest and slowest performance for each and the fastest remote
inference for DataScale. We observe that at mini-batch sizes below 1K, the
node-local RDU provides a lower latency than the A100. At mini-batch sizes in
the range $[4, 256]$ the measured latency of the remote inference on the
DataScale is lower than the latency of the most optimized node-local A100.  As
the mini-batch size increases above 256, the node-local performance of the A100
exceeds first remote and then node-local performance of the DataScale System.

From the measured latency of inference on the A100 and DataScale presented in
Figure~\ref{fig:sn_gpu_compare_latency}, we find that neither GPU or dataflow
architectures dominate the performance landscape. However, for the CogSim
application of the Hermit model, small mini-batch size performance is most
important. Therefore, we observe that our disaggregated system out-performs
traditional GPUs for our specific \texttt{in-the-loop} inference.

More generally, we observe the Hermit model at smaller mini-batch sizes is more
performant on the DataScale System while larger mini-batch sizes are faster
on the A100. Given the different behaviors of optimized configurations for
Hermit and MIR models in Section~\ref{sec:eval_gpu}, we find that the viability
of disaggregated systems for CogSim workloads is heavily driven by choice of
surrogate model and latency tolerances.

\subsection{Throughput tests}
\label{sec:eval_throughput}

\begin{figure}[!htb]
\centering
 \includegraphics[width=\columnwidth]{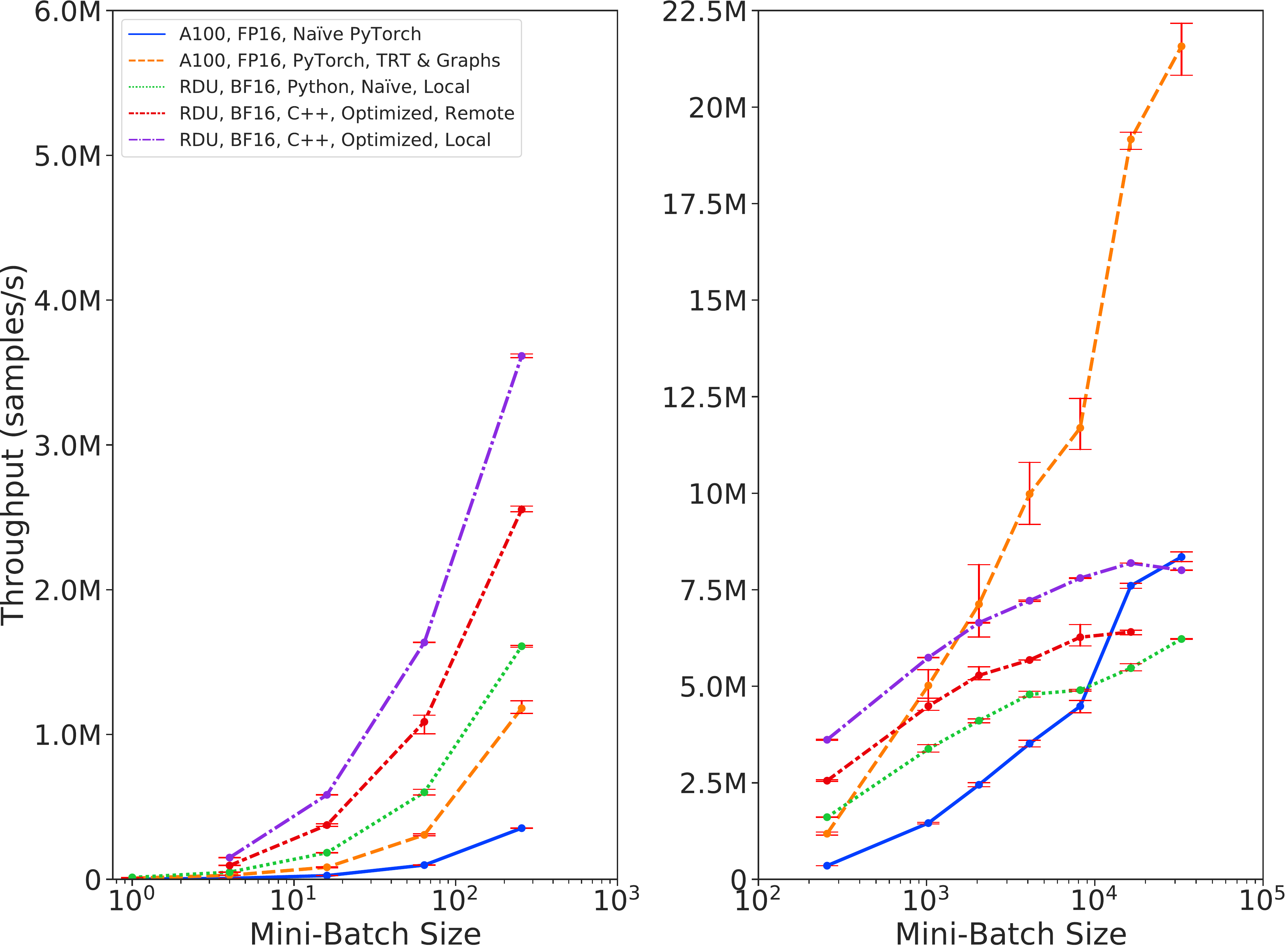}
 \caption{Inference throughput of the Hermit model on 1 RDU and A100 with
various configurations.}
 \label{fig:sn_gpu_compare_throughput}
\end{figure}

Throughput tolerances are also important for CogSim workloads.
Figure~\ref{fig:sn_gpu_compare_throughput} shows the Hermit inference
throughput for the same configurations of the A100 and DataScale as in
Figure~\ref{fig:sn_gpu_compare_latency}. In the left panel of the figure, we
see that in all configurations at mini-batch sizes below 1K, the DataScale is
measured to have the largest throughput. As the mini-batch size increases above
1K, the A100 throughput exceeds the DataScale throughput. We also observe
remote inference has less bandwidth than the node-local inference on a single
RDU. This drop in throughput bandwidth is a result of additional latency and
overhead associated with the high-speed network connecting the remote client to
the DataScale System.

\begin{figure}[!htb]
\centering
 \includegraphics[width=\columnwidth]{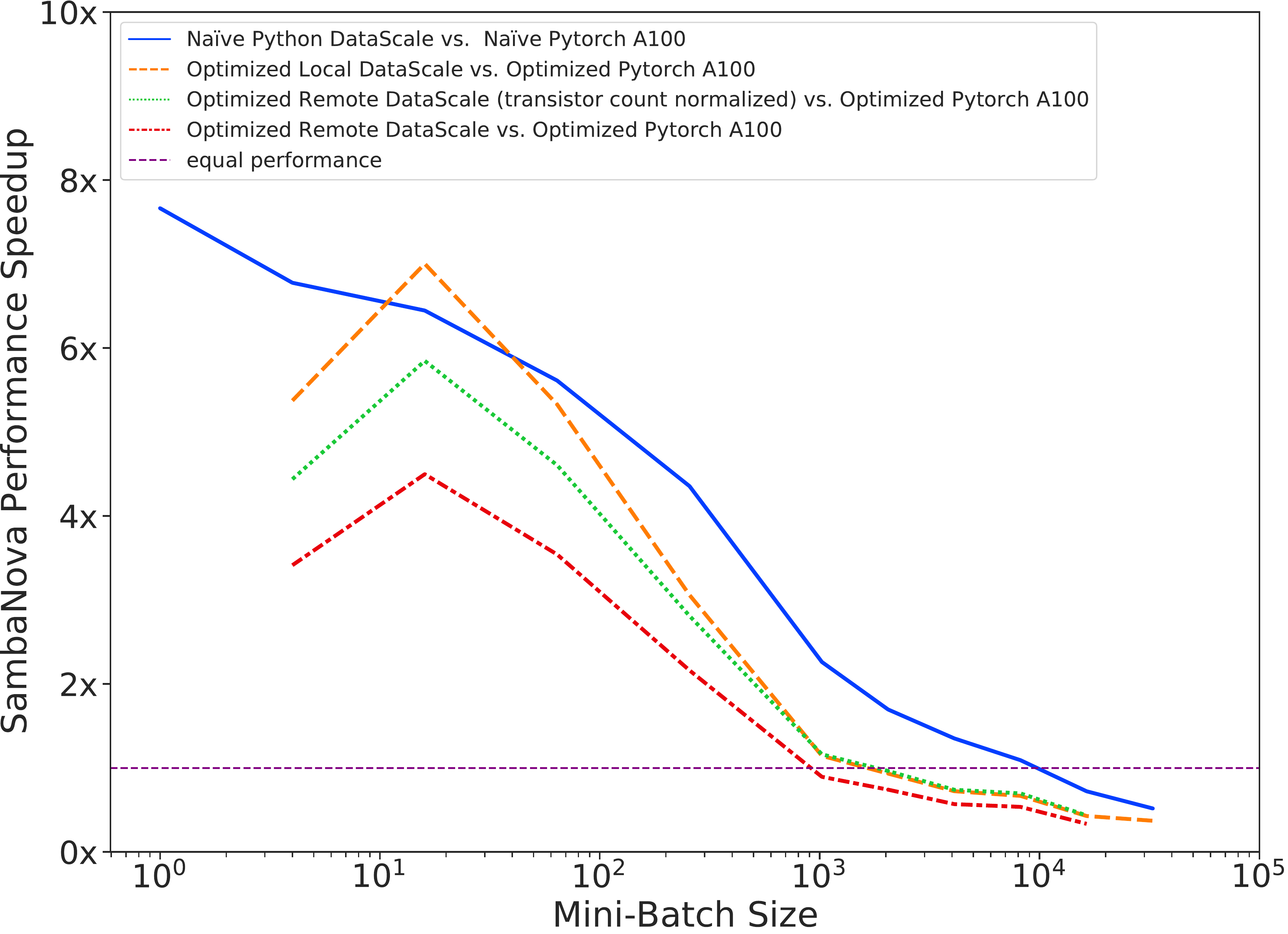}
 \caption{The inference latency speedup on 1 RDU compared to the A100 under
various configurations.}
 \label{fig:sn_gpu_compare_speedup}
\end{figure}

The relative performance measured on a single RDU of the DataScale compared to
the A100 for different configurations with the Hermit model is shown in
Figure~\ref{fig:sn_gpu_compare_speedup}. This plot shows the speedup factor of
the DataScale System on the Y-axis across all tested mini-batch sizes. We
compare 3 sets of configurations between the A100 and RDU: (1) na\"{i}ve
PyTorch implementations (i.e., the slowest), (2) optimized node-local
implementations (i.e., the fastest), and (3) optimized A100 node-local and
optimized DataScale remote (i.e., CogSim workload). We include another set of
datapoints for configuration (3) where we normalize the DataScale throughput by
transistor count. The A100 has 1.3x the transistor count of the DataScale RDU.
Above the horizontal dotted line of Figure~\ref{fig:sn_gpu_compare_speedup}
indicates higher throughput bandwidth on the DataScale compared to the A100.
These comparisons show that the DataScale dominates performance at the lower
mini-batch sizes with the Hermit model. The largest difference in performance
is between the most highly optimized node-local measurements, with a more than
7X speedup. In the context of CogSim and \texttt{in-the-loop} inference, we
find that the remote inference DataScale measured performance is more than 3X
that of the most highly optimized node-local A100 for the smallest mini-batch
sizes.  As the mini-batch sizes increase above 1K, the DataScale System lags
behind the A100. At the largest mini-batch sizes, the A100 offers better
performance.

\begin{figure}[!htb]
\centering
 \includegraphics[width=\columnwidth]{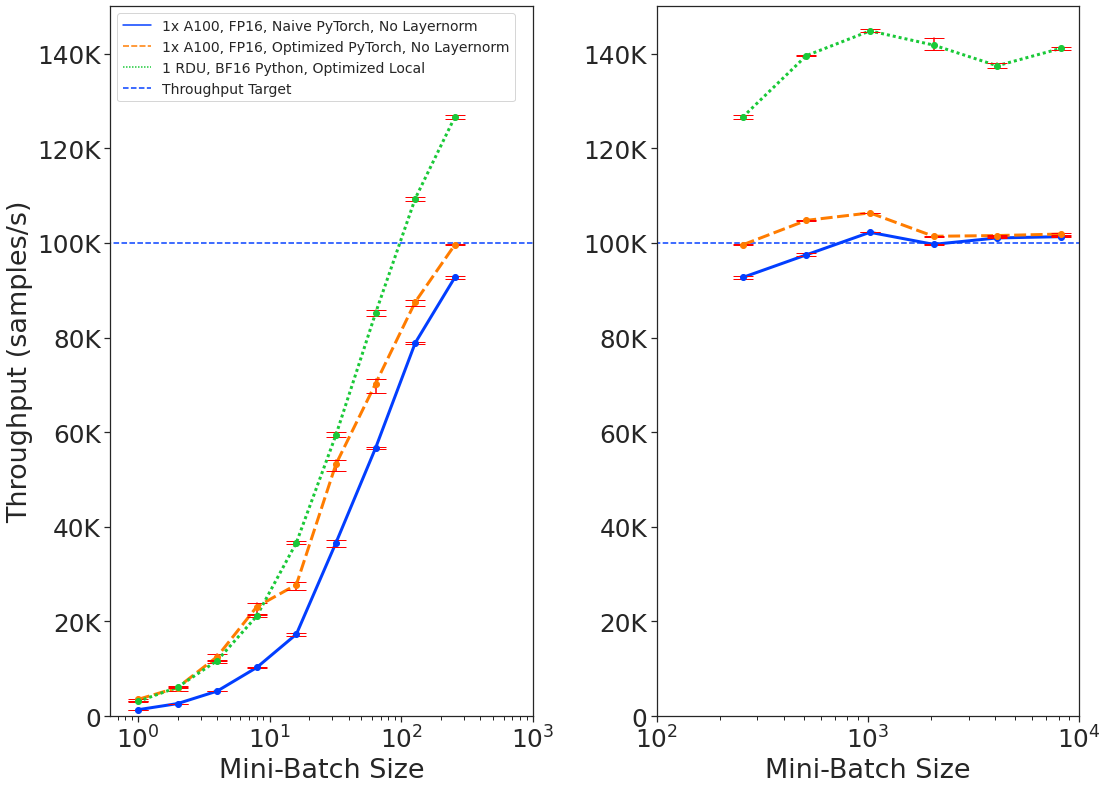}
 \caption{Inference throughput of the MIR model on 1 RDU and A100 with
different configurations.}
 \label{fig:sn_gpu_compare_mir}
\end{figure}

Switching to the MIR model, which has a target throughput of at least 100K
samples/s, we compare DataScale and Nvidia GPU performance in
Figure~\ref{fig:sn_gpu_compare_mir}. This comparison is done on a version of
the MIR model without layernorm to ensure the model would compile optimally on
both architectures. In this Figure, we show the target throughput with a
horizontal dashed line. We observe that at low mini-batch sizes, throughput is
similar between the A100 and DataScale. The DataScale system reaches the target
throughput bandwidth at a mini-batch size of 128 while the A100 reaches it at
size 256. As the mini-batch size increases toward 8K, the DataScale system
reaches a maximum throughput of over 140K while the A100 struggles to achieve a
throughput much larger than 100K. This result is contrasting to the results
with Hermit in
Figures~\ref{fig:sn_gpu_compare_throughput}~and~\ref{fig:sn_gpu_compare_speedup},
where the DataScale provided the largest advantage over the A100 at small
mini-batch sizes.

Our comparisons of measured latency in Section~\ref{sec:eval_latency} and
throughput in this section for the DataScale and A100 reveal that new dataflow
architectures are viable for disaggregated CogSim workloads. Specifically, we
demonstrated that with the Hermit model at mini-batch sizes below 1K and with
the MIR model at large mini-batch sizes the DataScale dominates performance.
This indicates that disaggregated systems are not only competitive, but
potentially faster than traditional GPU accelerators in these contexts.  From
differences and similarities in the performance landscapes between the two
tested models, we conclude that the DataScale and A100 are more performant
based on several contributing factors, including model size and
throughput/latency requirements.  For example, in the context of tight latency
requirements for large mini-batches with Hermit, the A100 would be the best
option. Our current results are model specific and cannot capture all the
complexities of the contributing factors. We discuss how our results can be
generalize to better demonstrate the viability of disaggregated systems in the
following section.

\section{Conclusion}

In this paper, we explored the viability of disaggregated systems for CogSim
workloads, including \texttt{in-the-loop} inference. We tested our hypothesis using two
surrogate models, Hermit and MIR, across three Nvidia GPUs, two AMD GPUs, and the new
SambaNova DataScale system. Measuring inference latency and throughput across
various configurations of the hardware, we described the performance
landscape and determining factors of DL performance. Our results indicate that
disaggregated systems are viable and in some cases offer better performance
than node-local GPU accelerators for CogSim workloads. We found that the
optimal hardware for CogSim workloads is largely determined by the model and
latency/throughput requirements.

Throughout this work we engaged very closely with vendor engineering teams to
ensure that our applications were mapped as efficiently as possible for each
accelerator architecture.  We found that that there were opportunities to use
these example applications to identify areas for improving the vendor's tool
chains for CogSim SciML workflows.  One area that is the subject of ongoing
work is a generalized application for remote inference on the DataScale, which
supports remote inference to multiple, independent models that is necessary for
the Hermit and MIR integration.

While our findings are relevant to CogSim workloads and more specifically
applications that use Hermit and MIR models for \texttt{in-the-loop} inference, our
results are not extensive enough to generalize to all workloads. The results we
obtained indicate that factors like model size, layer types, computational
requirements of the simulation, optimizability of the model, as well as
inference latency and throughput requirements are contributing factors to
determining the viability of a disaggregated system for a particular workload.
Our future work aims to explore this space by extending our results to
more automatically generated DL models that represent a wide array of
CogSim applications. This work would serve as a reference for other researchers
to indicate if a disaggregated system is viable for a given CogSim application.

\section*{Acknowledgment}
We thank Adam Moody (LLNL) for his contributions to this work. This work was
performed under the auspices of the U.S. Department of Energy by Lawrence
Livermore National Laboratory. LLNL-CONF-826438.
\par

\bibliographystyle{IEEEtran}
\bibliography{references}

\end{document}